\documentclass[conference]{IEEEtran}

\usepackage[pdftex]{graphicx}
\usepackage{array}
\usepackage{url}
\usepackage{microtype}
\usepackage{textcomp}
\usepackage{listings}
\usepackage[usenames,dvipsnames,svgnames,table]{xcolor}
\usepackage{amssymb}
\usepackage{booktabs}
\usepackage{makecell}
\usepackage{multirow}
\usepackage{ifthen}
\usepackage{tikz}
\usetikzlibrary{fit}
\usepackage{float}
\usepackage{pgfplots}
\usepackage{pgfplotstable}
\usepackage{hyperref}
\usepackage{balance}
\usepackage{cite} 

\makeatletter
\def\blfootnote{\xdef\@thefnmark{}\@footnotetext}
\makeatother

\usepackage{xspace}
\usepackage{relsize}
\def\ifmonospace{\ifdim\fontdimen3\font=0pt }
\def\C++{%
\ifmonospace%
    \C++%
\else%
    C\kern-.1167em\raise.30ex\hbox{\smaller{++}}%
\fi%
\spacefactor1000 }

\newcommand{\vthead}[1]{\makebox[1em][l]{\rotatebox{90}{#1}}}

\newcommand{\code}[1]{\texttt{\small#1}}
\usepackage{pifont}
\newcommand{\checkm}{\ding{52}}

\newcommand{\pie}[1]{
  \hspace{-.8ex}\hskip-\tabcolsep
  \begin{tikzpicture}[scale=0.8, baseline=-.5ex]
    \draw (0,0) circle (1ex);
    \fill[white] (0,0) circle (1ex);
    \ifthenelse{\equal{#1}{0}}{
    }{
      \fill (0,1ex) arc (90:int(#1*90+90):1ex) -- (0,0) -- cycle;
    }
  \end{tikzpicture}
  \hspace{-1.6ex}\hskip-\tabcolsep
}
\newcommand{\piev}[1]{
  \hspace{-.25ex}\hskip-\tabcolsep
  \begin{tikzpicture}[scale=0.8, baseline=-.5ex]
    \node[fit={(.75ex,-.4ex) (1.7ex,1.4ex)}, inner sep=0] {{\tiny \ding{91}}};
    \draw (0,0) circle (1ex);
    \fill[white] (0,0) circle (1ex);
    \ifthenelse{\equal{#1}{0}}{
    }{
      \fill (0,1ex) arc (90:int(#1*90+90):1ex) -- (0,0) -- cycle;
    }
  \end{tikzpicture}
  \hspace{-1.6ex}\hskip-\tabcolsep
}
\newcommand{\piel}[1]{\hspace{3pt}\pie{#1}\hspace{5pt}}
\newcommand{\pievl}[1]{\hspace{.12em}\piev{4}\hspace{5pt}}
\newcommand{\celll}{
  \hspace{-1.3ex}
  \begin{tikzpicture}[scale=0.8, baseline=.4ex]
    \fill[red!35] (0,0) rectangle (2.5ex,2.5ex);
  \end{tikzpicture}
  \hspace{-0.6ex}
}
\newcommand{\na}{{\scriptsize{n/a}}}
\newcommand{\fp}{{\cellcolor{red!35}}}

\begin{document}

\title{SoK: Sanitizing for Security}

\author{\IEEEauthorblockN{
    Dokyung Song,
    Julian Lettner,
    Prabhu Rajasekaran,\\
    Yeoul Na,
    Stijn Volckaert,
    Per Larsen,
    Michael Franz}
  \IEEEauthorblockA{
    University of California, Irvine\\
    {\tt \{dokyungs,jlettner,rajasekp,yeouln,stijnv,perl,franz\}@uci.edu}
    }
}

\maketitle

\blfootnote{\textcopyright{} 2018 IEEE. Personal use of this material is
  permitted. Permission from IEEE must be obtained for all other uses, in any
  current or future media, including reprinting/republishing this material for
  advertising or promotional purposes, creating new collective works, for resale
  or redistribution to servers or lists, or reuse of any copyrighted component
  of this work in other works.}

\begin{abstract}
The  C  and \C++  programming  languages  are  notoriously insecure  yet  remain
indispensable.  Developers therefore resort to  a multi-pronged approach to find
security issues  before adversaries. These  include manual, static,  and dynamic
program analysis.   Dynamic bug finding  tools---henceforth ``sanitizers''---can
find bugs  that elude other  types of analysis  because they observe  the actual
execution of  a program,  and can therefore  directly observe  incorrect program
behavior as it happens.

A vast number of sanitizers have been prototyped by academics and refined by
practitioners. We provide a systematic overview of sanitizers with an
emphasis on their role in finding security issues. Specifically, we
taxonomize the available tools and the security vulnerabilities they cover,
describe their performance and compatibility properties, and highlight various
trade-offs.
\end{abstract}

\IEEEpeerreviewmaketitle

\section{Introduction}
\label{sec:motivation}

C and \C++ remain the languages of choice for low-level systems software such as
operating system kernels, runtime libraries, and browsers. A key reason is that
they are efficient and leave the programmer in full control of the underlying
hardware. On the flip side, the programmer must ensure that every memory access
is valid, that no computation leads to undefined behavior, etc.  In practice,
programmers routinely fall short of meeting these responsibilities and introduce
bugs that make the code vulnerable to exploitation.

At the same time, memory corruption exploits are getting more
sophisticated~\cite{Shacham2007, carlini2015controlflowbending,
  schuster.etal+15, hu2016dop}, bypassing widely-deployed mitigations such as
Address Space Layout Randomization (ASLR) and Data Execution Prevention
(DEP). Code-reuse attacks such as Return-Oriented Programming (ROP) corrupt
control data such as function pointers or return addresses to hijack the
control-flow of the program~\cite{Shacham2007}.  Data-only attacks such as
Data-Oriented Programming (DOP) leverage instructions that can be invoked on
legal control-flow paths to compromise a program by corrupting only its
non-control data~\cite{hu2016dop}.

As a first line of defense against bugs, programmers use analysis tools to
identify security problems before their software is deployed in production.
These tools rely on either static program analysis, dynamic program analysis, or
a combination.
Static tools analyze the program source code and produce results that are
conservatively correct for all possible executions of the
code~\cite{clarke2004tool,godefroid2005dart,sen2005cute,cadar2008exe,cadar2008klee}.
In contrast, dynamic bug finding tools---often called ``sanitizers''---analyze a
single program execution and output a precise analysis result valid for a single
run only.

Sanitizers are now in widespread use and responsible for many vulnerability
discoveries. However, despite their ubiquity and critical role in finding
vulnerabilities, sanitizers are often not well-understood, which hinders their
further adoption and development. This paper provides a systematic overview of
sanitizers with an emphasis on their role in finding security
vulnerabilities. We taxonomize the available tools and the security
vulnerabilities they cover, describe their performance and compatibility
properties, and highlight various trade-offs. To foster further adoption and
development, we conclude the paper with viable deployment directions for
developers and future research directions for researchers.

\subsection{Exploit Mitigations vs. Sanitizers}
\label{sec:motivation:mitigations_vs_sanitizers}

\begin{table}
  \caption{Exploit Mitigations vs. Sanitizers}
  \vspace{-0.5em}
  \centering
  \label{tab:mitigate_sanitize}
  \resizebox{0.48\textwidth}{!}{
  \begin{tabular}{ l l l }
    \toprule
    & Exploit Mitigations & Sanitizers \\
    \midrule
    The goal is to ... & Mitigate attacks & Find vulnerabilities \\
    Used in ... & Production & Pre-release \\
    Performance budget ... & Very limited & Much higher \\
    Policy violations lead to ... & Program termination & Problem diagnosis \\
    Violations triggered at location of bug ... & Sometimes & Always \\
    Tolerance for FPs is ... & Zero & Somewhat higher \\
    Surviving benign errors is ... & Desired & Not desired \\
    \bottomrule
  \end{tabular}
  }
  \vspace{-1.5em}
\end{table}

Sanitizers are similar to many well-known exploit mitigations in that both types
of tools insert inlined reference monitors (IRMs) into the program to enforce a
fine-grained security policy. Despite this similarity, however, exploit
mitigations and sanitizers significantly differ in what they aim to achieve and
how they are used. These differences reflect in the design requirements, as
shown in Table~\ref{tab:mitigate_sanitize}.

The biggest difference between the two types of tools lies in the type of
security policy they enforce. Exploit mitigations deploy a policy aimed at
detecting or preventing attacks, whereas sanitizers aim to pinpoint the precise
locations of buggy program statements. Control-Flow Integrity
(CFI)~\cite{abadi2005cfi}, Data-Flow Integrity
(DFI)~\cite{castro2006dataflowintegrity} and Write Integrity Testing
(WIT)~\cite{akritidis2008preventing} are examples of exploit mitigations because
they detect deviations from legal control or data flow paths, which usually
happen as a consequence of a bug's exploitation, but do not necessarily happen
at the precise locations of vulnerable program statements. Bounds checking
tools, in contrast, could be considered sanitizers because violations of their
policies trigger directly at the locations of vulnerable statements.

Exploit mitigations are meant to be deployed in production, thus put stringent
requirements on various design aspects. First, exploit mitigations rarely see
real-world adoption if they incur non-negligible run-time
overhead~\cite{szekeres.etal+13}. Sanitizers have less stringent performance
requirements because they are only used for testing. Second, false positive
detections in an exploit mitigations are unacceptable because they terminate the
program. Sanitizers may tolerate false alerts to the extent that developers are
willing to review false bug reports.  Finally, surviving benign errors (e.g.,
writes to padding) is allowed and often desired in production systems for
reliability and availability reasons, whereas sanitizers aim to detect these
bugs precisely since their exploitability is unknown.

\section{Low-level Vulnerabilities} \label{sec:bugs}

Among a wide spectrum of security-related bugs, this paper covers bugs that
have specific security implications in C/\C++. This includes not only undefined
behavior, but also well-defined behaviors that are potentially dangerous in the
absence of type and memory safety. We briefly describe the bugs and how they can
be exploited to leak information, escalate privilege, or execute arbitrary code.

\subsection{Memory Safety Violations} \label{sec:bugs:memorysafety}

A program is \emph{memory safe} if pointers in the program only access their
\emph{intended referents}, while those intended referents are valid.
The intended referent of a pointer is the object from whose base address the
pointer was derived. Depending on the type of the referent, it is either valid
between its allocation and deallocation (for heap-allocated referents), between
a function call and its return (for stack-allocated referents), between the
creation and the destruction of its associated thread (for thread-local
referents), or indefinitely (for global referents).

Memory safety violations are among the most severe security vulnerabilities and
have been studied extensively in the
literature~\cite{szekeres.etal+13,van2012memory}. Their exploitation can lead to
code injection~\cite{stacksmashing}, control-flow
hijacking~\cite{SolarDesigner1997,Nergal2001,Shacham2007}, privilege
escalation~\cite{dataonly}, information leakage~\cite{strackx2009breaking}, and
program crashes.

\subsubsection{Spatial Safety Violations}  \label{sec:bugs:memorysafety:spatial}

Accessing memory that is not (entirely) within the bounds of the intended
referent of a pointer constitutes a spatial safety violation.
Buffer overflows are a typical example of a spatial safety violation. A buffer
overflow happens when the program writes beyond the end of a buffer.
If the intended referent of a vulnerable access is a subobject (e.g., a struct
field), and if an attacker writes to another subobject within the same object,
then we refer to this as an \emph{intra-object overflow}.
\autoref{lst:oob} shows an intra-object overflow vulnerability which can be
exploited to perform a privilege escalation attack.

\begin{lstlisting}[language=C, morekeywords={bool}, caption=Intra-object overflow vulnerability which can be exploited to overwrite security-critical non-control data, label=lst:oob]
struct A { char name[7]; bool isAdmin; };
struct A a; char buf[8];
memcpy(/* dst */ a.name, /* src */ buf, sizeof(buf));
\end{lstlisting}

\subsubsection{Temporal Safety Violations}  \label{sec:bugs:memorysafety:temporal}

A temporal safety violation occurs when the program accesses a referent that is
no longer valid. When an object becomes invalid, which usually happens by
explicitly deallocating it, all the pointers pointing to that object become
\emph{dangling}. Accessing an object through a dangling pointer is called a
\emph{use-after-free}.
Accessing a local object outside of its scope or after the function returns is
referred to as \emph{use-after-scope} and \emph{use-after-return},
respectively. This type of bug becomes exploitable when the attacker can reuse
and control the freed region, as illustrated in \autoref{lst:uaf}.

\begin{lstlisting}[language=C, morekeywords={}, caption=Use-after-free vulnerability which can be exploited to hijack the control-flow of the program, label=lst:uaf]
struct A { void (*func)(void); };
struct A *p = (struct A *)malloc(sizeof(struct A));
free(p);   // Pointer becomes dangling
...
p->func(); // Use-after-free
\end{lstlisting}

\vspace{-0.5em}
\subsection{Use of Uninitialized Variables} \label{sec:bugs:uum}

Variables have an \emph{indeterminate value} until they are
initialized~\cite{stdc11, stdcpp14}. \C++14 allows this indeterminate value to
propagate to other variables if both the source and destination variables have
an unsigned narrow character type. Any other use of an uninitialized variable
results in undefined behavior.
\noindent The effects of this undefined behavior depend on many factors,
including the compiler and compiler flags that were used to compile the
program. In most cases, indeterminate values are in fact the (partial) contents
of previously deallocated variables that occupied the same memory range as the
uninitialized variable. As these previously deallocated variables may sometimes
hold security-sensitive values, reads of uninitialized memory may be part of an
information leakage attack, as illustrated in \autoref{lst:uum:leakage}.

\begin{lstlisting}[language=C, morekeywords={}, caption=Use of a partially-initialized variable which becomes vulnerable as the uninitialized value crosses a trust boundary, label=lst:uum:leakage]
struct A { int data[2]; };
struct A *p = (struct A *)malloc(sizeof(struct A));
p->data[0] = 0; // Partial initialization
send_to_untrusted_client(p, sizeof(struct A));
\end{lstlisting}

\vspace{-0.5em}
\subsection{Pointer Type Errors} \label{sec:bugs:pointertype}

C and \C++ support several casting operators and language constructs that can
lead memory accesses to misinterpret the data stored in their referents, thereby
violating type safety.
Pointer type errors typically result from unsafe casts. C allows all casts
between pointer types, as well as casts between integer and pointer types. The
\C++ \code{reinterpret\_cast} type conversion operator is similarly not subject
to any restrictions. The \code{static\_cast} and \code{dynamic\_cast} operators
do have restrictions. \code{static\_cast} forbids pointer to integer casts,
and casting between pointers to objects that are unrelated by
inheritance. However, it does allow casting of a pointer from a base class to a
derived class (also called \emph{downcasting}), as well as all casts from and to
the \code{void*} type.
\emph{Bad-casting} (often referred to as \emph{type confusion}) happens when a
downcast pointer has neither the run-time type of its referent, nor one of the
referent's ancestor types.

\begin{lstlisting}[language=C++, morekeywords={}, caption=Bad-casting vulnerability leading to a type- and memory-unsafe memory access, label=lst:badcasting]
class Base { virtual void func(); };
class Derived : public Base { public: int extra; };
Base b[2];
Derived *d = static_cast<Derived *>(&b[0]); // Bad-casting
d->extra = ...; // Type-unsafe, out-of-bounds access, which
                // overwrites the vtable pointer of b[1]
\end{lstlisting}

\noindent To downcast safely, programmers must use the \code{dynamic\_cast}
operator, which performs run-time type checks and returns a null pointer if the
check fails. Using \code{dynamic\_cast} is entirely optional, however, and
introduces additional run-time overhead.

Type errors can also occur when casting between function pointer types. Again,
\C++'s \code{reinterpret\_cast} and C impose no restrictions on casts between
incompatible function pointer types. If a function is called indirectly through
a function pointer of the wrong type, the target function might misinterpret its
arguments, which leads to even more type errors.
Finally, C also allows type punning through \code{union} types. If the program
reads from a union through a different member object than the one that was used
to store the data, the underlying memory may be misinterpreted. Furthermore, if
the member object used for reading is larger than the member object used to
store the data, then the upper bytes read from the union will take unspecified
values.

\subsection{Variadic Function Misuse} \label{sec:bugs:variadic}

C/\C++ support \emph{variadic functions}, which accept a variable number of
\emph{variadic} function arguments in addition to a fixed number of regular
function arguments. The variadic function's source code does not specify the
number or types of these variadic arguments. Instead, the fixed arguments and the
function semantics encode the expected number and types of variadic arguments.
Variadic arguments can be accessed and simultaneously typecast using \code{va\_arg}.
It is, in general, impossible to statically verify that \code{va\_arg} accesses a
valid argument, or that it casts the argument to a valid type. This lack of
static verification can lead to type errors, spatial memory safety violations,
and uses of uninitialized values.

\begin{lstlisting}[language=C++, morekeywords={}, caption=Simplified version of CVE-2012-0809; user-provided input was mistakenly used as part of a larger format string passed to a printf-like function, label=lst:fmt]
char *fmt2; // User-controlled format string
sprintf(fmt2, user_input, ...);
// prints attacker-chosen stack contents if fmt2 contains
// too many format specifiers
// or overwrites memory if fmt2 contains %n
printf(fmt2, ...);
\end{lstlisting}

\subsection{Other Vulnerabilities} \label{sec:bugs:others}

There are other operations that may pose security risks in the absence of type
and memory safety. Notable examples include overflow errors which may be
exploitable when such values are used in memory allocation or pointer arithmetic
operations. If an attacker-controlled integer value is used to calculate a
buffer size or an array index, the attacker could overflow that value to
allocate a smaller buffer than expected (as illustrated in
Listing~\ref{lst:signedintegeroverflow}), or to bypass existing array index
checks, thereby triggering an out-of-bounds access.

\begin{lstlisting}[language=C, morekeywords={}, caption=Simplified version of CVE-2017-5029; a signed integer overflow vulnerability that can lead to spatial memory safety violation, label=lst:signedintegeroverflow]
// newsize can overflow depending on len
int newsize = oldsize + len + 100;
newsize *= 2;
// The new buffer may be smaller than len
buf = xmlRealloc(buf, newsize);
memcpy(buf + oldsize, string, len); // Out-of-bounds access
\end{lstlisting}

\noindent C/\C++ do not define the result of a signed integer overflow, but
stipulate that unsigned integer wrap around when they overflow. However, this
wrap-around behavior is often unintended and potentially dangerous.

Undefined behaviors such as signed integer overflows pose additional security
risks when compiler optimizations are enabled.  In the presence of potential
undefined behavior, compilers are allowed to assume that the program will never
reach the conditions under which this undefined behavior is triggered. The
compiler can then generate optimized program code based on this
assumption~\cite{lee2017undefllvm}. Consequently, the compiler does not have to
statically verify that the program is free of potential undefined behavior, and
the compiler does not need to generate code that is capable of recognizing and
mitigating undefined behavior.  The problem with this rationale is that
optimizations based on the assumption that the program is free from undefined
behavior can sometimes lead the compiler to omit security checks. In
CVE-2009-1897, for example, GCC infamously omitted a null pointer check from one
of the Linux kernel drivers, which led to a privilege escalation
vulnerability~\cite{null-pointer-deref-exploit}. Compiler developers regularly
add such aggressive optimizations to their compilers. Some people therefore
refer to potential undefined behavior as \emph{time
  bombs}~\cite{dietz2012understanding}.

\begin{lstlisting}[language=C++, morekeywords={}, caption=Simplified version of CVE-2009-1897; dereferencing a pointer lets the compiler safely assume that the pointer is non-null, label=lst:optnullchk]
struct sock *sk = tun->sk; // Compiler assumes tun is not
                           // a null pointer
if (!tun) // Check is optimized out
  return POLLERR;
\end{lstlisting}

\section{Bug Finding Techniques} \label{sec:techniques}

We now review the relevant bug finding techniques. We begin each subsection with
an informal description of the bug finding policy, followed by a description of
mechanisms that implement (or approximate) that policy.

\subsection{Memory Safety Violations} \label{sec:sanitizers:memorysafety}

Memory safety bug finding tools detect dereferences of pointers that either do
not target their intended referent (i.e., spatial safety violations), or that
target a referent that is no longer valid (i.e., temporal safety
violations). There are two types of tools for detecting these bugs. We summarize
their high-level goals and properties here, and then proceed with an in-depth
discussion of the techniques these tools can employ to detect memory safety
bugs.

\paragraph*{\textbf{Location-based Access Checkers}}

Location-based access checkers detect memory accesses to invalid memory regions.
These checkers have a metadata store that maintains state for each byte of (a
portion) of the addressable address space, and consult this metadata store
whenever the program attempts to access memory to determine whether the memory
access is valid or not.
Location-based access checkers can use red-zone
insertion~\cite{hastings1991purify, seward2005memcheck, bruening2011drmem,
  hasabnis2012lbc, serebryany2012addresssanitizer} or guard
pages~\cite{perens1993electric,microsoft2000pageheap} to detect spatial safety
violations. Either of these techniques can be combined with memory-reuse delay
to additionally detect temporal safety
violations~\cite{hastings1991purify,jones1997bounds,seward2005memcheck,bruening2011drmem,serebryany2012addresssanitizer,perens1993electric,microsoft2000pageheap,dhurjati2006dangling,dang2017oscar}.
Location-based access checkers incur low run-time performance overheads, and are
highly compatible with non-instrumented code.
The downside is that these tools are imprecise, as they can only detect if an
instruction accesses valid memory, but \emph{not if the accessed memory is part
  of the intended referent of the instruction}. These tools generally incur high
memory overhead.

\paragraph*{\textbf{Identity-based Access Checkers}}

Identity-based access checkers detect memory accesses that do not target their
intended referent. These tools maintain metadata (e.g., bounds or allocation
status) for each allocated memory object, and have a mechanism in place to
determine the intended referent for every pointer in the program. Metadata
lookups can happen when the program calculates a new pointer using arithmetic
operations to determine if the calculation yields a valid pointer and/or upon
pointer dereferences to determine if the dereference accesses the intended
referent of the pointer.
Identity-based access checkers can use per-object bounds
tracking~\cite{jones1997bounds, ruwase2004cred, dhurjati2006bounds,
  akritidis2009baggy, eigler2003mudflap, younan2010paricheck,
  duck2016lowfatheap, duck2017lowfatstack} or per-pointer bounds
tracking~\cite{kendall1983bcc, steffen1992adding, austin1994efficient,
  patil1997low, nagarakatte2009softbound, intel-pointerchecker, necula2002ccured,
  jim2002cyclone, xu2004mscc, nethercote2004bounds} to detect spatial safety
violations, and can be extended with lock-and-key
checking~\cite{austin1994efficient, patil1997low, nagarakatte2010cets} or with
dangling pointer
tagging~\cite{caballero2012undangle,lee2015dangnull,younan2015freesentry,kouwe2017dangsan}
to detect temporal safety violations.
Identity-based checkers are more precise than location-based access checkers, as
they cannot just detect accesses to invalid memory, but also accesses to valid
memory outside of the intended referent. 
These tools do, however, incur higher run-time performance overhead than
location-based checkers. Identity-based checkers are generally not compatible
with non-instrumented code. They also have higher false positive detection rates
than location-based checkers.

\subsubsection{Spatial Memory Safety Violations} \label{sec:sanitizers:memorysafety:spatial}

\paragraph*{\textbf{Red-zone Insertion}}  \label{sec:sanitizers:memorysafety:spatial:red-zone}

Location-based access checkers can insert so-called \emph{red-zones} between
memory objects~\cite{hastings1991purify, seward2005memcheck, bruening2011drmem,
  hasabnis2012lbc, serebryany2012addresssanitizer}.  These red-zones represent
out-of-bounds memory and are marked as invalid memory in the metadata store.
Memory regions that are part of allocated memory pages, but that are not part of
any static or dynamic allocation in the program are similarly considered to be
red-zones and are therefore marked as invalid in the metadata store. Any access
to a red-zone or to an unallocated memory region triggers a warning.
Purify was the first tool to employ this
technique~\cite{hastings1991purify}. Purify inserts the red-zones at the
beginning and the end of each allocation.
Purify tracks the state of the program's allocated address space using a large
shadow memory bitmap that stores two bits of state per byte of
memory. Valgrind's Memcheck uses the same technique but reserves two bits of
state for every bit of memory~\cite{seward2005memcheck}. Consequently, Memcheck
can detect access errors with bit-level precision, rather than byte-level
precision.

Light-weight Bounds Checking (LBC) similarly inserts red-zones, but adds a fast
path to the location-based access checks to reduce the overhead of the metadata
lookups~\cite{hasabnis2012lbc}.  LBC does this by filling the red-zones with a
random pattern and assumes compares the data read/overwritten by every memory
access with the fill pattern. If the data does not match the fill pattern, the
access is considered safe because it could not have targeted a red-zone. If the
data does happen to match the fill pattern, LBC performs a secondary slow path
check that looks up the state of the accessed data in the metadata store, and
triggers a warning if the accessed data is a red-zone.

Location-based access checkers that use red-zone insertion generally incur low
run-time performance overhead, but have limited precision as they can only
detect illegal accesses that target a red-zone. Illegal accesses that target a
valid object, which may or may not be part of the same allocation as the
intended referent, cannot be detected. Red-zone insertion-based tools also fail
to detect intra-object overflow bugs because they do not insert red-zones
between subobjects. While technically feasible, inserting red-zones between
subobjects would lead to excessive memory overhead and it would change the layout of
the parent object. Any code that accesses the parent object or one of its
subobjects would therefore have to be modified, which would also break
compatibility with external code that is not aware of the altered data layout.

\paragraph*{\textbf{Guard Pages}} \label{sec:sanitizers:memorysafety:spatial:guardpages}

Location-based access checkers can insert inaccessible guard pages before and/or
after every allocated memory
object~\cite{perens1993electric,microsoft2000pageheap}.  Out-of-bound reads and
writes that access a guard page trigger a page fault, which in turn triggers an
exception in the application.  This use of the paging hardware to detect illegal
accesses allows location-based access checkers to run without instrumenting
individual load and store instructions. Using guard pages does, however, incur
high memory overhead, making the technique impractical for applications with
large working sets. Microsoft recognized this problem and added an option to
surround memory objects with guard blocks instead of full guard pages in
PageHeap~\cite{microsoft2000pageheap}. PageHeap fills these guard blocks with a
fill pattern, and verifies that the pattern is still present when a memory
object is freed. This technique is strictly inferior to red-zone insertion, as
it only detects out-of-bounds writes (and not reads), and it does not detect the
illegal writes until the overwritten object is freed.

\paragraph*{\textbf{Per-pointer Bounds Tracking}}  \label{sec:sanitizers:memorysafety:spatial:pointer}

Identity-based access checkers can store bounds metadata for every
pointer~\cite{kendall1983bcc, steffen1992adding, austin1994efficient,
  patil1997low, nagarakatte2009softbound, intel-pointerchecker, necula2002ccured,
  jim2002cyclone, xu2004mscc, nethercote2004bounds}. Whenever the program
creates a pointer by calling \code{malloc} or by taking the address of an
object, the tracker stores the base and size of the referent as metadata for the
new pointer. The tracker propagates this metadata when the program calculates
new pointers through arithmetic and/or assignment operations. Systems such as
Bcc~\cite{kendall1983bcc} instrument array accesses and output a warning if the
program reads or writes outside the bounds of an array. Subsequent tools such as
Safe-C~\cite{austin1994efficient} and SoftBound~\cite{nagarakatte2009softbound}
instrument all pointer dereferences and warn when a pointer is outside the
bounds of its intended referent when the program attempts to dereference it.

Identity-based access checkers that use per-pointer bounds tracking can provide
complete spatial memory violation detection, including detection of intra-object
overflows. SoftBound~\cite{nagarakatte2009softbound} and Intel Pointer
Checker~\cite{intel-pointerchecker} detect intra-object overflows by narrowing
the pointer bounds to the bounds of the subobject whenever the program derives a
pointer from the address of a subobject (i.e., a struct field).  The primary
disadvantage of per-pointer bounds tracking is poor compatibility, as the program
generally cannot pass pointers to non-instrumented libraries because such
libraries do not propagate or update bounds information correctly.  Another
disadvantage is that per-pointer metadata propagation adds high run-time
overheads.  CCured reduces this overhead by identifying ``safe'' pointers, which
can be excluded from bounds checking and metadata
propagation~\cite{necula2002ccured}. However, even with such optimizations, per-pointer
bounds checking remains expensive without hardware
support~\cite{devietti2008hardbound}.

\paragraph*{\textbf{Per-object Bounds Tracking}} \label{sec:sanitizers:memorysafety:spatial:object}

Identity-based access checkers can also store bounds metadata for every memory
object, rather than for every pointer~\cite{jones1997bounds, ruwase2004cred,
  dhurjati2006bounds, akritidis2009baggy, eigler2003mudflap,
  younan2010paricheck, duck2016lowfatheap, duck2017lowfatstack}.

This approach---pioneered by Jones and Kelly (J\&K)---solves some of the
compatibility issues associated with per-pointer bounds
tracking~\cite{jones1997bounds}. Per-object bounds trackers can maintain bounds
metadata without instrumenting pointer creation and assignment operations. The
tracker only needs to intercept calls to memory allocation (i.e., \code{malloc})
and deallocation (i.e., \code{free}) functions, which is possible even in
programs that are not fully instrumented.  Since bounds metadata is maintained
only for objects and not for pointers, it is difficult to link
pointers to their intended referent. While the intended referent of an in-bounds
pointer can be found using a range-based lookup in the metadata store, such a
lookup would not return the correct metadata for an out-of-bounds (OOB) pointer.
J\&K therefore proposed to instrument pointer arithmetic operations, and to
invalidate pointers as they go out-of-bounds (OOB). Any subsequent dereference
triggers a fault, which can then be caught to output a warning.

J\&K's approach, however, does not allow OOB pointers to be converted back into
in-bounds pointers. CRED does allow such conversions if the resulting pointer
points back at the original referent~\cite{ruwase2004cred}. To do so, CRED links
OOB pointers to so-called \emph{OOB objects} which store the address of the
original referent of the OOB pointer.

Baggy Bounds Checking (BBC) eliminates the need to allocate dedicated OOB
objects by storing the distance between the OOB pointer and its referent into
the pointer's most significant bits~\cite{akritidis2009baggy}. Tagging the most
significant bits also turns OOB pointers into invalid user-space pointers, such
that dereferencing them causes a fault. BBC compresses the size of the per-object
metadata by rounding up all allocation sizes to the nearest power of two,
such that one byte of metadata suffices to store the bounds.

Low-fat pointer (LFP) bounds checkers improve BBC by embedding the allocation
size in the pointer representation without turning them into invalid user-space
pointers~\cite{duck2016lowfatheap, duck2017lowfatstack}. Pointers can therefore
be passed to non-instrumented code which does not know how to remove the tag.
The idea behind LFP is to partition the heap into equally-sized subheaps that
each supports only one allocation size. Thus, the allocation size for any given
pointer can be obtained by looking up the allocation size supported by that
heap. The base address can be calculated by rounding it down to the allocation
size. LFP allows pointers to go OOB and to be converted back into in-bounds
pointers only if the OOB pointer does not escape the current context. To ensure
that all escaping pointers are within bounds, LFP inserts a bounds check for
pointer arithmetic whose resulting pointer can be passed to a different context.

Per-object bounds trackers can support near-complete spatial safety
vulnerability detection. Allocator-based techniques such as BBC and LFP,
however, do sacrifice precision for better run-time performance as they check
allocation bounds rather than object bounds (cf.
Section~\ref{sec:analysis:falsenegatives}).

Per-object bounds tracking has other downsides too. First, per-object bounds
trackers do not detect intra-object overflows (cf.
Section~\ref{sec:analysis:falsenegatives}). Second, marking pointers as OOB by
pointing them to an OOB object, or by writing tags into their upper bits might
impact compatibility with external code that is unaware of the bounds checking
scheme used in the program. Specifically, external code is unable to restore OOB
pointers into in-bounds pointers even when that is the intent.

\subsubsection{Temporal Memory Safety Violations} \label{sec:sanitizers:memorysafety:temporal}

\paragraph*{\textbf{Memory-reuse Delay}} \label{sec:sanitizers:memorysafety:temporal:delay}

Location-based access checkers can mark recently deallocated objects as invalid
in the metadata store by replacing them with
red-zones~\cite{hastings1991purify,jones1997bounds,seward2005memcheck,bruening2011drmem,serebryany2012addresssanitizer}
or with guard
pages~\cite{perens1993electric,microsoft2000pageheap,dhurjati2006dangling,dang2017oscar}.
The existing location-based checking mechanism can then detect dangling pointer
dereferences as long as the deallocated memory is not reused. If the program
does reuse deallocated memory to allocate new objects, this approach will
erroneously allow dangling pointer dereferences to proceed. Some memory-reuse
delay-based tools reduce the chance of such detection failures by delaying the
reuse of recently deallocated memory regions until they have
``aged''~\cite{hastings1991purify,jones1997bounds,seward2005memcheck,bruening2011drmem,serebryany2012addresssanitizer}. This
leads to a trade-off between precision and memory overhead as longer reuse
delays lead to higher memory overhead, but also to a higher chance of detecting
dangling pointer dereferences.

Dhurjati and Adve (D\&A) proposed to use static analysis to determine exactly
when deallocated memory is safe to reuse~\cite{dhurjati2006dangling}. D\&A
allocate every memory object on its own virtual memory page, but allow objects
to share physical memory pages. This is possible by aliasing virtual memory
pages to the same physical page. When the program frees a memory object, D\&A
convert its virtual page into a guard page. D\&A also partition the heap into
pools, leveraging a static analysis called Automatic Pool
Allocation~\cite{lattner2005poolalloc}. This analysis can infer when a pool is
no longer accessible (even through potentially dangling pointers), at which
point all virtual pages in the pool can be reclaimed.  Dang et al. proposed a
similar system that does not use pool allocation, and can therefore be applied
to source-less programs~\cite{dang2017oscar}. Similar to D\&A, Dang et
al. allocate all memory objects on their own virtual pages. Upon deallocation of
an object, Dang et al. unmap that object's virtual page. This effectively
achieves the same goal as guard pages, but allows the kernel to free additional
page table entries, which reduces the physical memory overhead. To prevent reuse
of unmapped virtual pages, Dang et al. propose to map new pages at the high
water mark (i.e., the highest virtual address that has been used in the
program). While this does not rule out reuse of unmapped virtual pages
completely, the idea is that reuse is unlikely to happen given a 64-bit address space.

\paragraph*{\textbf{Lock-and-key}} \label{sec:sanitizers:memorysafety:temporal:allocid}

Identity-based checkers can detect temporal safety violations by assigning
unique allocation identifiers---often called keys---to every allocated
memory object and by storing this key in a \emph{lock
  location}~\cite{austin1994efficient, patil1997low, nagarakatte2010cets}. They
also store the lock location and the expected key as metadata for every
pointer. The checker revokes the key from the lock location when its associated
object is deallocated. Lock-and-key checking detects temporal memory safety
violations when the program dereferences a pointer whose key does not match the
key stored in the lock location for that pointer. Assuming unique keys, this
approach provides complete coverage of temporal safety
violations~\cite{nagarakatte2010cets}. Since this technique stores per-pointer
metadata, it naturally complements identity-based checkers that also detect
spatial violations using per-pointer bounds tracking. The drawbacks of
lock-and-key checking are largely the same as those for per-pointer bounds
tracking: compatibility with non-instrumented code is poor because
non-instrumented code will not propagate the metadata correctly, and the
run-time overhead is significant because maintaining metadata for every pointer
is expensive.

\paragraph*{\textbf{Dangling Pointer Tagging}} \label{sec:sanitizers:memorysafety:temporal:danglingpointertracking}

The most straightforward way to tag dangling pointers is to nullify either the
value or the bounds associated with pointers that are passed to the \code{free}
function~\cite{intel-pointerchecker}. A spatial memory safety violation
detection mechanism would then trigger a warning if such pointers are
dereferenced at a later point in time. The disadvantage of this approach is that
it does not tag copies of the dangling pointer, which may also be used in an
unsafe way.

Several tools tag not only the pointer passed to \code{free}, but also copies of
that pointer by maintaining auxiliary data structures that link all memory
objects to any pointers that refer to
them~\cite{caballero2012undangle,lee2015dangnull,younan2015freesentry,kouwe2017dangsan}.
Undangle uses taint
tracking~\cite{newsome2005dynamic,suh2004secure,chow2004understanding} to track
pointer creations, and to maintain an object-to-pointer
map~\cite{caballero2012undangle}. Whenever the program deallocates a memory
object, Undangle can query this pointer map to quickly find all dangling
pointers to the now deallocated object. Undangle aims to report not only the use
but also the existence of dangling pointers. It has a configurable time window
where it considers dangling pointers latent but not unsafe, e.g., a transient
dangling pointer that appears during the deallocation of nested
objects. Undangle reports a dangling pointer when this window expires, or
earlier if the program attempts to dereference the pointer.

DangNull~\cite{lee2015dangnull}, FreeSentry~\cite{younan2015freesentry}, and
DangSan~\cite{kouwe2017dangsan} steer clear of taint tracking and instrument all
pointer creations at compile time instead. These tools maintain pointer maps by
calling a runtime registration function whenever the program assigns a
pointer. Whenever the program deallocates a memory object, the tools look up all
pointers that point to the object being deallocated, and invalidate
them. Subsequent dereferences of an invalidated dangling pointer result in a
hardware trap.

Dangling pointer tagging tools that are not based on taint tracking have
some fundamental limitations. First, they require the availability of
source as it relies on precise type information to determine which operations
store new pointers. Second, they fail to maintain accurate metadata if
the program copies pointers in a type-unsafe manner (e.g., by casting them to
integers). Third, and most importantly, they can only link objects to
pointers stored in memory, and is therefore unaware of dangling pointers stored
in registers. Taint tracking-based tools such as Undangle, have none of these
disadvantages, but incur significant performance and memory overheads.

\subsection{Use of Uninitialized Variables} \label{sec:sanitizers:uum}

These tools detect uses of uninitialized values.

\paragraph*{\textbf{Uninitialized Memory Read Detection}}

Location-based access checkers can be extended to detect reads of uninitialized
memory values by marking all memory regions occupied by newly allocated objects
as uninitialized in the metadata store~\cite{hastings1991purify}. These tools
instrument read instructions to trigger a warning if they read from
uninitialized memory regions, and they instrument writes to clear the
uninitialized flag for the overwritten region. Note that marking memory regions
as uninitialized is not equivalent to marking them as a red-zone, since both
read and write accesses to red-zones should trigger a warning, whereas accesses
to uninitialized memory should only be disallowed for reads.

\paragraph*{\textbf{Uninitialized Value Use Detection}}

Detecting reads of uninitialized memory yields many false positive detections,
as the \C++14 standard explicitly allows uninitialized values to propagate
through the program as long as they are not used. This happens, for example,
when copying a partially uninitialized struct from one location to the other.
Memcheck attempts to detect only the uses of uninitialized values by limiting
error reporting to (i) dereferences of pointers that are (partially) undefined,
(ii) branching on a (partially) undefined value, (iii) passing undefined values
to system calls, and (iv) copying uninitialized values into floating point
registers~\cite{seward2005memcheck}. To support this policy, Memcheck adds one
byte of shadow state for every partially initialized byte in the program
memory. This allows Memcheck to track the definedness of all of the program's
memory with bit-level precision. Memcheck approximates the \C++14 semantics but
produces false negatives (failing to report illegitimate uses of uninitialized
memory) and false positives (reporting legitimate uses of uninitialized memory),
which are unavoidable given that Memcheck operates at the binary level, rather
than the source code level.
MemorySanitizer (MSan) implements fundamentally the same policy as Memcheck, but
instruments programs at the compiler Intermediate Representation (IR)
level~\cite{stepanov2015memorysanitizer}. The IR code carries more information
than binary code, which makes MSan more precise than Memcheck. MSan produces no
false positives (provided that the entire program is instrumented) and few false
negatives. Its performance overhead is also an order of magnitude lower than
Memcheck.

\subsection{Pointer Type Errors} \label{sec:sanitizers:pointertype}

These tools detect bad-casting or dereferencing of pointers of incompatible types.

\paragraph*{\textbf{Pointer Casting Monitor}} Pointer casting monitors detect
illegal downcasts through the \C++ \code{static\_cast} operator. Illegal
downcasts occur when the target type of the cast is not equal to the run-time
type (or one of its ancestor types) of the source object.
UndefinedBehaviorSanitizer~\cite{ubsan} (UBSan) and Clang CFI~\cite{clang-cfi}
include checkers that verify the correctness of \code{static\_cast} operations by comparing the target
type to the run-time type information (RTTI) associated with the source object. This
effectively turns \code{static\_cast} operations into \code{dynamic\_cast}.
The downside is that RTTI-based tools cannot verify casts
between non-polymorphic types that lack RTTI.

CaVer~\cite{lee2015caver} and TypeSan~\cite{haller2016typesan} do not rely on
RTTI to track type information, but instead maintain custom metadata for all
types and all objects used in the program. This way, they can extend the
type-checking coverage to non-polymorphic types. At compile time, these tools
build per-class type metadata tables which contain all the valid type casts for
a given pointer type.  The type tables encode the class inheritance
relationships and information on how classes with multiple base classes are
composed (i.e., at which offset within objects of a derived class is each base
object located).
Both tools also track the effective run-time types for each live object by
monitoring memory allocations and storing the allocated types in a metadata
store. To perform downcast checking, the tools retrieve the run-time type of the
source object from the metadata store, and then query the type table for the
corresponding class to check if the type conversion is in the table (and is
therefore permissible). HexType similarly tracks type information in disjoint
metadata structures, but provides a more accurate run-time type
tracing~\cite{jeon2017hextype}. HexType also replaces
the default implementation of \code{dynamic\_cast} with its own
optimized implementation, while preserving its run-time semantics, i.e.,
returning NULL for failing casts.

\paragraph*{\textbf{Pointer Use Monitor}}

C/\C++ support several constructs to convert types in potentially dangerous
ways; C-style casts, \code{reinterpret\_cast}, and unions can all be used to
bypass compile-time type checking. Extending pointer casting monitoring to all
three of these constructs can result in false positives, however, as
type confusion bugs can only be exploited if a pointer with the wrong type gets
dereferenced. For this reason, one might opt for pointer dereference/use
monitoring over pointer casting monitoring.

Loginov et al. proposed a pointer use monitor for C
programs~\cite{loginov2001debugging}.
The tool maintains and verifies run-time type tags for each memory location by
monitoring load and store operations. A tag contains the scalar type that was
last used to write to its corresponding memory location. Aggregate types are
supported by breaking them down into their scalar components. The tool stores
the tags in shadow memory.
Whenever a value is read from memory, the tool checks if the type used to load
the value matches the type tag.
A more recent pointer use monitor is TypeSanitizer (TySan)~\cite{tysan}, which
is being integrated into the LLVM project as of this writing.
TySan leverages type information generated by the compiler frontend (clang) to
maintain a type tag store in shadow memory, and to verify the correctness of
load instructions. Contrary to Loginov et al.'s tool, however, TySan does not
require that the types used to store and load from a memory location match
exactly. Instead, TySan only requires type compatibility, as defined by the
aliasing rule in the C/\C++ standard.
For example, TySan allows all loads through a character pointer type, even if
the target location was stored using a pointer to a larger type. Loginov et
al.'s tool would detect this as an error, but this behavior is explicitly
permitted by the language standard. TySan also differs from Loginov et al.'s
tool in that it checks struct types at a finer granularity. Loginov et al. build
type tags for structs by fully decomposing them into their scalar
components. \code{struct A\{int a;\};} and \code{struct A\{struct B\{int a;\}
  b;\};} would therefore get the same type tag. TySan, by contrast, is struct
path-aware and can distinguish these two types.

Several tools also detect pointer type errors in indirect function calls, that
is, calling functions through a pointer of a type incompatible with the type of
the callee~\cite{ubsan,parasoft-insure,clang-cfi}. Function-signature-based
forward-edge control flow integrity mechanisms such as Clang
CFI~\cite{clang-cfi} can be viewed as sanitizers that detect such function
pointer misuses. Since all the function signatures are known at compile-time,
these tools can detect mismatches between the pointer type and the function type
without maintaining run-time tags.

\subsection{Variadic Function Misuse} \label{sec:sanitizers:variadic}

These tools detect memory safety violations and uninitialized variable uses
specific to variadic functions.

\paragraph*{\textbf{Dangerous Format String Detection}}

The most prominent class of variadic function misuse bugs are format string
vulnerabilities. Most efforts therefore focus solely on detection of dangerous
calls to \code{printf}. Among these efforts are tools that restrict the usage of
the \code{\%n} qualifier in the format
string~\cite{tsai2001libsafe,ringenburg2005preventing}. This qualifier may be
used to have \code{printf} write to a caller-specified location in
memory. However, this dangerous operation~\cite{carlini2015controlflowbending}
is specific to the \code{printf} function, so the aforementioned tools'
applicability is limited.

\paragraph*{\textbf{Argument Mismatch Detection}}

FormatGuard prevents \code{printf} from reading more arguments than were passed
by the caller~\cite{cowan2001formatguard}. FormatGuard does this by redirecting
the calls to a protected \code{printf} implementation that increments a counter
whenever it retrieves a variadic argument through \code{va\_arg}. If the counter
surpasses the number of arguments specified at the call site, FormatGuard raises
an alert. HexVASAN generalizes argument counting to all variadic functions, and
also adds type checking~\cite{biswas2017vasan}. HexVASAN instruments the call
sites of variadic functions to capture the number and types of arguments passed
to the callee and saves this information in a metadata store. The tool then
instruments \code{va\_start} and \code{va\_copy} operations to retrieve
information from the metadata store, and it instruments \code{va\_arg}
operations to check if the argument being accessed is within the given number of
arguments and of the given type.

\subsection{Other Vulnerabilities}

These tools detect other undefined behavior or dangerous well-defined but
potentially unintended behavior.

\paragraph*{\textbf{Stateless Monitoring}}
UndefinedBehaviorSanitizer (UBSan) is~\cite{ubsan}, to our knowledge, the only
dynamic tool that detects any of the types of undefined behavior we have not
covered so far.  The undefined behaviors UBSan detects currently include signed
integer overflows, floating point or integer division by zero, invalid bitwise
shift operations, floating point overflows caused by casting (e.g., casting a
large double-precision floating point number to a single-precision float), uses
of misaligned pointers, performing an arithmetic operation that overflows a
pointer, dereferencing a null pointer, and reaching the end of a value-returning
function without returning a value.  Most of UBSan's detection features are
stateless, so they can be turned on collectively without interfering with each
other.  UBSan can also detect several kinds of well-defined but likely
unintended behavior. For example, the language standard dictates that unsigned
integers have their value reset to zero when they overflow. This well-defined
behavior is often unexpected and a frequent source of bugs, however, so UBSan
can optionally detect these unsigned integer wraparounds.

\section{Program Instrumentation} \label{sec:sanitizers:embedding} \label{sec:transformation}

Sanitizers implement their bug finding policy by embedding inlined reference
monitors (IRMs) into the program. These IRMs monitor
and mediate all program instructions that can contribute to a
vulnerability. Such instructions include (but are not limited to) memory loads
and stores, stack frame (de)allocations, calls to memory allocation functions
(e.g., \code{malloc}), and system calls. IRMs can be embedded using a compiler,
linker, or an instrumentation framework.

\subsection{Language-level Instrumentation} \label{sec:instr:language-level}
Sanitizers can be embedded at the source code or abstract syntax tree (AST)
level.  The source code and AST are language-specific and typically contain full
type information, language-specific syntax, and compile time-evaluated
expressions such as \code{const\_cast} and \code{static\_cast} type conversions.
This language-specific information is typically discarded when the compiler
lowers the AST into Intermediate Representation (IR) code. Language-level
instrumentation is recommended (or even necessary) for sanitizers that detect
pointer type errors through pointer cast monitoring.

An additional advantage of instrumenting at the language level is that the
compiler preserves the full semantics of the program throughout the early stages
of the compilation. The sanitizer can therefore see the semantics intended by
the programmer. At later stages in the compilation, the compiler may assume that
the program contains no undefined behavior, and it may optimize the code based
on this assumption (e.g., by eliminating seemingly unnecessary security
checks). The disadvantage of instrumenting at the language level is that the
entire source code of the application must be available and the code must be
written in the expected language. Thus, this approach does not work for
applications that link against closed-source libraries, nor does it work for
applications that contain inline assembly code.

\subsection{IR-level Instrumentation}

Sanitizers can also be embedded later stage in the compilation, when the AST has
been lowered into IR code. Compiler backends such as LLVM support IR-level
instrumentation~\cite{LLVM}.  This approach is more generic than source-level
transformation in that the compiler IR is (mostly) independent of the source
language. Thus, by instrumenting at this level, the sanitizer can automatically
support multiple source languages.  An additional advantage is that the compiler
backend implements various static analyses and optimization passes that can be
used by the sanitizer. Sanitizers can leverage this infrastructure to optimize
the IRMs they embed into the program (e.g., by removing redundant or provably
safe checks).

The disadvantage of IR-level instrumentation is largely similar to that of
language-level instrumentation, i.e., the lack of support for closed-source
libraries and inline assembly code (Section~\ref{sec:instr:language-level}).
Exceptionally, ASan does provide limited support for inline x86 assembly
code at the IR level because it includes a pass that is dedicated to parsing and
instrumenting \code{MOV} and \code{MOVAPS} instructions found in inline assembly
blocks~\cite{serebryany2012addresssanitizer}.
Manual assembly parsers are architecture-specific, however, and need to be
reimplemented or duplicated for every supported architecture.

\subsection{Binary Instrumentation} \label{sec:instr:dyn:bin}
Dynamic binary translation (DBT) frameworks allow instrumentation of a program
at run time~\cite{bruening2003dynamorio, luk2005pin, nethercote2007valgrind}.
They read program code, instrument it, and translate it to machine code while
the program executes and expose various hooks to influence execution.  The main
advantage of DBT-based tools over compiler-based tools is that they work well
for closed-source programs.  Moreover, DBT frameworks offer complete
instrumentation coverage of user-mode code regardless of its origin. DBT
frameworks can instrument the program itself, third party code (that may be
dynamically loaded), and even dynamically generated code.

The main disadvantage of DBT is the (much) higher run-time performance overhead
compared to static instrumentation tools (see Section~\ref{sec:cost:perf}). This
overhead can be primarily attributed to run-time decoding and translation of
instructions. This problem can be partially addressed by instrumenting binaries
statically using a Static Binary Instrumentation (SBI) framework.  However, both
SBI and DBT-based sanitizers must operate on binaries that contain virtually no
type information or language-specific syntax. It is therefore impossible to
embed a pointer type error sanitizer at this stage. Information about stack
frame and global data section layouts is also lost at the binary level, which
makes it impossible to insert a fully precise spatial memory safety sanitizer
using binary instrumentation.

\subsection{Library Interposition} \label{sec:instr:dylib}

An alternative, albeit very coarse-grained, method is to intercept calls to
library functions using library interposers~\cite{curry1994profiling}. A library
interposer is a shared library that, when preloaded into a
program~\cite{preload}, can intercept, monitor, and manipulate all inter-library
function calls in the program. Some sanitizers use this method to intercept
calls to memory allocation functions such as \code{malloc} and \code{free}.

The advantage of this approach is that, similarly to DBT-based instrumentation,
it works well for COTS binaries in that no source or object code is required.
Contrary to DBT, however, library interposition incurs virtually no
overhead. One disadvantage is that library interposition only works for
\emph{inter-}library calls. A call between two functions in the same library
cannot be intercepted. Another disadvantage is that library interposition is
highly platform and target-specific. A sanitizer that uses library interposition
to intercept calls to \code{malloc} would not work for programs that have their
own memory allocator, for example.

\section{Metadata Management} \label{sec:metadata}

One important aspect of a sanitizer design is how it stores and looks up
metadata. This metadata typically captures information about the state of
pointers or memory objects used in the program. Although run-time performance is
not a primary concern for sanitizer developers or users, the sheer quantity of
metadata most sanitizers store means that even small inefficiencies in the
storage scheme can make the sanitizer unacceptably slow. The metadata storage
scheme also by and large determines whether two sanitizers can be used in
conjunction. If two independent sanitizers both use a metadata scheme that
changes the pointer and/or object representation in the program, they often
cannot be used together.

\subsection{Object Metadata}

Some sanitizers use object metadata storage schemes to store state for all
allocated memory objects. This state may include the object size, type, state
(e.g., allocated/deallocated, initialized/uninitialized), allocation identifier,
etc.

\paragraph*{\textbf{Embedded Metadata}} \label{sec:meta:obj:inline}
An obvious way to store metadata for an object is to increase its
allocation size and to append or prepend the metadata to the object's data. This
mechanism is popular among modern memory allocators which, for example, store
the length of a buffer in front of the actual buffer. Tools can modify memory
allocators to transparently reserve memory for metadata in addition to the
requested buffer size, and then return a pointer into the middle of this
allocation so that the metadata is invisible to clients. Allocation-embedded
metadata is used in ASan~\cite{serebryany2012addresssanitizer} and
Oscar~\cite{dang2017oscar}, among others. ASan embeds information about the
allocation context in each allocated object. Oscar stores each object's
canonical address as embedded metadata.

\paragraph*{\textbf{Direct-mapped Shadow}} \label{sec:meta:obj:direct}
The direct-mapped shadow scheme maps every block of $n$ bytes in the
application's memory to a block of $m$ bytes of metadata via the formula:

\begin{lstlisting}[language=C++, morekeywords={}]
// shadow_base is the base address of the metadata store
// block_addr is the address of the memory block
metadata_addr = shadow_base + m * (block_addr / n)
\end{lstlisting}

ASan~\cite{serebryany2012addresssanitizer}, for example, stores $1$ byte of
metadata for every $8$ bytes of application memory. In this case, the shadow
mapping formula can be simplified to:

\begin{lstlisting}[language=C++, morekeywords={}]
metadata_addr = shadow_base + (block_addr >> 3)
\end{lstlisting}

The direct-mapped shadow scheme is easy to implement and insert into an
application. It is generally also very efficient since only one memory read is
needed to retrieve the metadata for any given object. There are cases where it
can lead to poor run-time performance too, however, as it can worsen memory
fragmentation (and thus spatial locality) in already fragmented address
spaces. It is also wasteful in terms of allocated memory, since the shadow
memory area is contiguous and must be big enough to mirror all allocated memory
blocks (from the lowest virtual addresses to the highest).

\paragraph*{\textbf{Multi-level Shadow}} \label{sec:meta:obj:twolevel}
The multi-level shadow scheme can reduce the memory footprint of metadata store
by introducing additional layers of indirection in the form of directory
tables. These directory tables can store pointers to metadata tables or other
directory tables. Each metadata table directly mirrors a portion of the
application memory, similar to the direct-mapped shadow scheme. As a whole, the
multi-level shadow scheme resembles how modern operating systems implement page
tables. Having additional layers of indirection allows metadata stores to
allocate metadata tables on-demand. They only have to allocate the directory
tables themselves, and can defer allocation of the metadata tables until they
are needed. This is particularly useful for systems that have limited address
space (e.g., 32-bit systems), where sanitizers that implement direct-mapped
shadow schemes (e.g., ASan~\cite{serebryany2012addresssanitizer}) often exhaust
the available address space and cause program termination.

Tools that require per-object metadata (in contrast to per-byte metadata) can
use a \emph{variable-compression-ratio} multi-level shadow mapping scheme, where
the directory table maps variable-sized objects to constant-sized metadata.
This scheme can help the tools to optimize their shadow memory usage and
allocation-time performance~\cite{haller2016metalloc}.

The main characteristic of this scheme is that each metadata access requires
multiple memory accesses: one for each level of directory tables and another one
for the corresponding metadata table.  This significantly affects performance,
especially for tools that frequently look up metadata, e.g., a bounds checking
tool which requires metadata access for most memory accesses.
TypeSan~\cite{haller2016typesan}, for example, is a good fit for the two-level
variable-compression-ratio scheme, as the type metadata is per-object and
constant-sized and metadata lookup is infrequent.

\paragraph*{\textbf{Custom Data Structure}} \label{sec:meta:misc}
In addition to variations of the previously presented metadata schemes, some
tool authors have opted for a range of custom data structures and tool-specific
solutions to store metadata.  Bounds checkers such as J\&K, CRED, and D\&A
employ splay trees~\cite{jones1997bounds, ruwase2004cred,
  dhurjati2006bounds}. UBSan and CaVer use an additional hash table as a cache
to store the most recent results of type checking. DangNull utilizes a
thread-safe red-black tree to encode the relationship between
objects~\cite{lee2015dangnull}. Note that, when using a data structure without
support for concurrent access it must be protected by explicit locks in a
multi-threaded setting. For thread-local or stack variables, per-thread metadata
is also a choice, e.g., CaVer's per-thread red-black tree for stack and global
objects.

\subsection{Pointer Metadata} \label{sec:metadata:pointer}

\paragraph*{\textbf{Fat Pointers}} \label{sec:meta:ptr}
Some sanitizers replace standard machine pointers with fat pointers.  Fat
pointers are structures that contain the original pointer value, as well as
metadata associated with the original pointer. A fairly straightforward fat
pointer layout, used in many per-pointer bounds tracking tools is:

\begin{lstlisting}[language=C++, morekeywords={size_t}]
struct fat_pointer {
  void* value; // Original pointer value
  void* base;  // Base address of the intended referent
  size_t size; // Size of the referent
};
\end{lstlisting}

The primary advantage of using fat pointers is that they do not add much
additional cache pressure compared to regular pointers, and that they can store
an arbitrary amount of metadata. The disadvantages are that they require
extensive instrumentation of the program, they change the calling conventions
for functions that accept pointer arguments (fat pointers occupy more than one
register when passed as a function argument), and that they cannot be used in
programs that interact with non-instrumented third-party libraries. Without
instrumentation, these libraries do not interpret the fat pointer correctly, nor
are they able to update the fat pointer when its embedded inner pointer value
changes.

\paragraph*{\textbf{Tagged Pointers}}

A less invasive way to store per-pointer metadata is to replace regular machine
pointers with tagged pointers. A tagged pointer embeds metadata in the pointer
itself, without changing its size. This technique provides better compatibility
than fat pointers. Passing tagged pointers as function arguments does not
require changes to the standard calling conventions, for example. Another
advantage is that tagged pointers do not introduce any additional cache pressure
compared to regular machine pointers.  The amount of information that a tagged
pointer can encode is limited by size of the unused address space for a given
target platform. Most AMD64 platforms, including Linux/x86\_64, for example,
only use the lowest $256$TiB of virtual address space for user-mode
applications. The upper $16$ bits of any valid user-mode pointer are therefore
guaranteed to be zero. These $16$ bits can store per-pointer metadata. Baggy
Bounds Checking~\cite{akritidis2009baggy} uses the spare bits to store the
distance between an OOB pointer and its intended referent.  Note that tagged
pointers usually cannot be dereferenced directly. The metadata should be masked
out whenever a tagged pointer is dereferenced. This also means that, similar to
fat pointers, tagged pointers must be unpacked if they escape to external
non-instrumented libraries. This is not true for low-fat pointer (LFP) bounds
checkers, which store the metadata implicitly in the tagged pointer value, and
can be dereferenced directly~\cite{duck2016lowfatheap, duck2017lowfatstack}.

\paragraph*{\textbf{Disjoint Metadata}} \label{sec:meta:ptr:disjoint}

Storing metadata in a disjoint metadata store instead of embedding it in the
pointer representation improves compatibility over the aforementioned
approaches.  In contrast to per-object metadata, however, sanitizers usually do
not use direct-mapped shadow stores to maintain per-pointer metadata
(cf. Section~\ref{sec:meta:obj:direct}).  The portion of memory occupied by
pointers is usually small and the size of pointer metadata (e.g., bounds) tends
to exceed the size of the pointer itself, resulting in wasteful consumption of
address space.  For this reason, even at the cost of additional memory accesses,
tool authors have favored multi-level structures for maintaining per-pointer
metadata.  CETS~\cite{nagarakatte2010cets} utilizes a two-level lookup trie
(similar to page tables) using the pointer location as the key to store the
allocation identifier and the lock address of the referent.  Intel Pointer
Checker~\cite{intel-pointerchecker} and Intel MPX~\cite{intel-mpx} also use a
two-level structure to maintain pointer bounds.

The main disadvantage of disjoint metadata compared to in-pointer metadata is
that the sanitizer must explicitly propagate the metadata whenever the program
copies a pointer to a new memory location. If the program calls \code{memcpy} to
copy a data structure containing pointers, for example, then the sanitizer must
update the metadata store for the pointers in the target data structure. With
in-pointer metadata, by contrast, the metadata always travels with the pointer.

\subsection{Static Metadata} \label{sec:meta:static}

Some sanitizers require certain information discarded by the compiler to perform
their checks at run time. To make the required compile-time information
available at run time, these sanitizers usually embed static metadata into the
compiled program.  For example, bad-casting sanitizers create a type hierarchy
table at compile time to facilitate type casting checks at run time. HexVASAN, a
variadic function call sanitizer, builds static metadata for each variadic call
site to encode the number of arguments and their types. At run time, the
instrumented caller pushes the static metadata onto a custom stack, which is
used by the callee to check the validity of the supplied arguments.

\section{Driving a Sanitizer} \label{sec:sanitizers:drivers}

Dynamic analysis tools, including sanitizers, only detect bugs on code paths
that execute during testing. Increasing path coverage therefore increases bug
finding opportunities. Program execution can be driven by unit or integration
test suites, automated fuzzers, alpha and beta testers, or any combination
thereof.

Unit testing and integration testing are already best practices in software
engineering. Writing these tests has traditionally been a manual process. While
indispensable in general, using hand-written tests does have some drawbacks when
used to sanitize a program. First, manually written tests often focus on
positive testing using \emph{valid} inputs to check expected behavior.  Security
bugs are typically exploited by feeding the program \emph{invalid} inputs,
however.  Second, manually written tests hardly ever cover \emph{all} code
paths. Developers can use automated test case generators to alleviate this
problem~\cite{ali2010systematic,sen2005cute, godefroid2005dart,
  mcminn2004search}, but this option is generally only available when the full
source code is available.

A second option is to run a fuzzer on the program being sanitized~\cite{afl,
  libfuzzer, stephens2016driller, bohme2016aflfast, rawat2017vuzzer,
  bohme2017aflgo}.  Fuzzers are testing tools that run programs on automatically
generated inputs.  The main advantages of fuzzers are that (i) they perform
negative testing, because they tend to provide invalid inputs to the program and
(ii) they can run automatically once integrated into a project.  Fuzzers can
find security bugs relatively quickly, especially if the bugs are triggered on
code paths that are easily accessible.

A third option to find security bugs is to ship sanitization-enabled programs to
beta testers and to collect and transmit any sanitizer output back to the
developer. The main advantage here is that beta testers can distribute the
testing load, therefore allowing developers to locate bugs more quickly. One
disadvantage is that beta testers will inadvertently focus on testing the
program's main usage scenarios. Another disadvantage is that sanitizers can slow
down the program so much that it becomes unusable on consumer-grade machines,
thus reducing the chance that beta testers will thoroughly test the programs.

\section{Analysis}
\label{sec:analysis}

\definecolor{light-gray}{gray}{0.75}
\newcolumntype{?}{!{\color{light-gray}\vrule width 0.4pt}}

\begin{table*}[!t]
  \centering
  \setlength{\tabcolsep}{0.15em}
  \rotatebox{0}{
  {\rowcolors{3}{lightgray!20}{white}
  \begin{tabular}{| l | cc | ccc?cccc?cc?ccc?c | ccc?c | ccc?ccc?c | ccc?ccc?ccc ? cc |}
    \hline
      \multicolumn{1}{|c|}{\thead{Sanitizers}}
    & \multicolumn{2}{c|}{} 
    & \multicolumn{13}{c|}{\thead{\ref{sec:techniques}. Bug Finding Techniques}}
    & \multicolumn{4}{c|}{\thead{\ref{sec:transformation}. Instr.}}
    & \multicolumn{7}{c|}{\thead{\ref{sec:metadata}. Metadata Mgmt.}}
    & \multicolumn{11}{c|}{\thead{\ref{sec:analysis}. Analysis}}
    \\

    & \vthead{Year Published}
    & \vthead{Actively Maintained}

    & \vthead{Red-zone Insertion}
    & \vthead{Guard Pages}
    & \vthead{Memory-reuse Delay}
    & \vthead{Per-pointer Bounds Tracking}
    & \vthead{Per-object Bounds Tracking}
    & \vthead{Lock-and-key}
    & \vthead{Dangling Pointer Tagging}
    & \vthead{Uninit. Mem. Read Detection}
    & \vthead{Uninit. Value Use Detection}
    & \vthead{Pointer Casting Monitor}
    & \vthead{Pointer Use Monitor}
    & \vthead{Variadic Arg. Mismatch Detection}
    & \vthead{Stateless Monitoring}

    & \vthead{Language-level Instr.}
    & \vthead{IR-level Instr.}
    & \vthead{Binary Instr.}
    & \vthead{Library Interposition}

    & \vthead{Embedded Metadata}
    & \vthead{Direct-mapped Shadow}
    & \vthead{Multi-level Shadow}
    & \vthead{Custom Data Structure}
    & \vthead{Fat/Tagged Pointers}
    & \vthead{Disjoint Per-pointer Metadata}
    & \vthead{Static Metadata}

    & \vthead{Spatial Safety Violation~(\ref{sec:bugs:memorysafety:spatial})}
    & \vthead{Temporal Safety Violation~(\ref{sec:bugs:memorysafety:temporal})}
    & \vthead{Use of Uninit. Variables~(\ref{sec:bugs:uum})}
    & \vthead{Bad-casting~(\ref{sec:bugs:pointertype})}
    & \vthead{Load Type Mismatch~(\ref{sec:bugs:pointertype})}
    & \vthead{Func. Call Type Mismatch~(\ref{sec:bugs:pointertype})}
    & \vthead{Variadic Func. Misuse~(\ref{sec:bugs:variadic})}
    & \vthead{Signed Integer Overflow~(\ref{sec:bugs:others})}
    & \vthead{Other Undef. Behavior~(\ref{sec:bugs:others})}
    & \vthead{Performance Overhead}
    & \vthead{Memory Overhead}
    \\
    \hline

    Purify~\cite{hastings1991purify}
    & '91 &
    & \checkm && \checkm
    &&&
    &
    & \checkm &
    &&&
    &
    &&& \checkm &
    && \checkm &&&&&
    & \pie{1} & \pie{2} & \pie{4}\fp{} &&&&&&
    & \pie{3} & \pie{4} \\

    Memcheck~\cite{seward2005memcheck}
    & '05 & \checkm
    & \checkm && \checkm
    &&&
    &
    && \checkm
    &&&
    &
    &&& \checkm &
    &&& \checkm &&&&
    & \pie{1} & \pie{2} & \pie{2} &&&&&&
    & \piev{0} 
    & \na
    \\

    Dr. Memory~\cite{bruening2011drmem}
    & '11 & \checkm
    & \checkm && \checkm
    &&&
    &
    && \checkm
    &&&
    &
    &&& \checkm &
    &&& \checkm &&&&
    & \pie{1} & \pie{2} & \pie{2} &&&&&&
    & \pie{0} 
    & \na
    \\

    LBC~\cite{hasabnis2012lbc}
    & '12 & 
    & \checkm &&
    &&&
    &
    &&
    &&&
    &
    & \checkm &&& 
    &&& \checkm &&&& 
    & \pie{3} &&&&&&&&
    & \pie{3} 
    & \pie{2} 
    \\

    ASan~\cite{serebryany2012addresssanitizer}
    & '12 & \checkm 
    & \checkm && \checkm
    &&&
    &
    &&
    &&&
    &
    && \checkm &&
    && \checkm &&&&&
    & \pie{3} & \pie{2} &&&&&&&
    & \piev{3} 
    & \pie{2} 
    \\

    Electric Fence~\cite{perens1993electric}
    & '93 &
    && \checkm & \checkm
    &&&
    &
    &&
    &&&
    &
    &&&& \checkm
    &&&&&&&
    & \pie{1} & \pie{3} &&&&&&&
    & \na & \pie{0}
    \\

    PageHeap~\cite{microsoft2000pageheap} 
    & '00 &
    && \checkm & \checkm
    &&&
    &
    &&
    &&&
    &
    &&&& \checkm
    &&&&&&&
    & \pie{1} & \pie{3} &&&&&&&
    & \na & \pie{0}
    \\

    D\&A Dangling~\cite{dhurjati2006dangling}
    & '06 &
    && \checkm & \checkm
    &&&
    &
    &&
    &&&
    &
    && \checkm && 
    &&&&&&&
    && \pie{3} &&&&&&&
    & \pie{3}
    & \na
    \\

    Oscar~\cite{dang2017oscar}
    & '17 &
    && \checkm & \checkm
    &&&
    &
    &&
    &&&
    &
    &&&& \checkm
    & \checkm &&&&&&
    && \pie{3} &&&&&&&
    & \pie{3} 
    & \pie{4} 
    \\

    RTCC~\cite{steffen1992adding}
    & '92 &
    &&&
    & \checkm &&
    &
    &&
    &&&
    &
    & \checkm &&&
    &&&&& \checkm &&
    & \pie{3} &&&&&&&&
    & \pie{2}
    & \na
    \\

    Safe-C~\cite{austin1994efficient}
    & '94 &
    &&&
    & \checkm && \checkm
    &
    &&
    &&&
    &
    & \checkm &&& 
    &&&&& \checkm & \checkm & 
    & \pie{4}\fp{} & \pie{4}\fp{} &&&&&&&
    & \pie{2} 
    & \pie{4} 
    \\

    P\&F~\cite{patil1997low}
    & '97 &
    &&&
    & \checkm && \checkm
    &
    &&
    &&&
    &
    & \checkm &&& 
    &&&&&& \checkm & 
    & \pie{4}\fp{} & \pie{4}\fp{} &&&&&&&
    & \pie{2}
    & \pie{4}
    \\

    MSCC~\cite{xu2004mscc}
    & '04 &
    &&&
    & \checkm && \checkm
    &
    &&
    &&&
    &
    & \checkm &&& 
    &&&&&& \checkm & 
    & \pie{3} & \pie{3} &&&&&&&
    & \pie{3} 
    & \pie{4} 
    \\

    SoftBound+CETS~\cite{nagarakatte2009softbound,nagarakatte2010cets}
    & '10 &
    &&&
    & \checkm && \checkm
    &
    &&
    &&&
    &
    && \checkm &&
    &&&&&& \checkm &
    & \pie{4}\fp{} & \pie{4}\fp{} &&&&&&&
    & \pie{3}
    & \pie{4}
    \\
    Intel Pointer Checker~\cite{intel-pointerchecker}
    & '12 & \checkm
    &&&
    & \checkm &&
    & \checkm
    &&
    &&&
    &
    && \checkm &&
    &&&&&& \checkm &
    & \pie{4}\fp{} & \pie{1} &&&&&&& 
    & \pie{3} 
    & \pie{4} 
    \\

    J\&K~\cite{jones1997bounds}
    & '97 & 
    &&&
    && \checkm &
    &
    &&
    &&&
    &
    & \checkm &&& 
    &&&& \checkm &&&
    & \pie{3}\fp{} &&&&&&&&
    & \pie{2} 
    & \na
    \\

    CRED~\cite{ruwase2004cred}
    & '04 & 
    &&&
    && \checkm &
    &
    &&
    &&&
    &
    & \checkm &&&
    &&&& \checkm &&&
    & \pie{3}\fp{} &&&&&&&&
    & \pie{3}
    & \na
    \\


    D\&A Bounds~\cite{dhurjati2006bounds}
    & '06 &
    &&&
    && \checkm &
    &
    &&
    &&&
    &
    && \checkm && 
    &&&& \checkm &&&
    & \pie{3}\fp{} &&&&&&&&
    & \pie{3}
    & \na
    \\

    BBC~\cite{akritidis2009baggy}
    & '09 &
    &&&
    && \checkm &
    &
    &&
    &&&
    &
    && \checkm && 
    &&&&& \checkm &&
    & \pie{2}\fp{} & &&&&&&&
    & \pie{3}
    & \pie{4}
    \\

    PAriCheck~\cite{younan2010paricheck}
    & '10 &
    &&&
    && \checkm &
    &
    &&
    &&&
    &
    & \checkm &&&
    &&&& \checkm &&&
    & \pie{2}\fp{} &&&&&&&&
    & \pie{3}
    & \na
    \\

    Low-fat Pointer~\cite{duck2016lowfatheap, duck2017lowfatstack}
    & '17 &
    &&&
    && \checkm &
    &
    &&
    &&&
    &
    && \checkm && 
    &&&&& \checkm &&
    & \pie{2}\fp{} &&&&&&&& 
    & \piev{3} 
    & \pie{4} 
    \\

    Undangle~\cite{caballero2012undangle}
    & '12 &
    &&&
    &&&
    & \checkm
    &&
    &&&
    &
    &&& \checkm &
    &&&&&& \checkm &&
    & \pie{3} &&&&&&&
    & \pie{0}
    & \pie{0}
    \\

    FreeSentry~\cite{younan2015freesentry}
    & '15 &
    &&&
    &&&
    & \checkm
    &&
    &&&
    &
    & \checkm &&&
    &&& \checkm & \checkm &&&
    && \pie{3}\fp{} &&&&&&& 
    & \pie{3} 
    & \na
    \\

    DangNull~\cite{lee2015dangnull}
    & '15 &
    &&&
    &&&
    & \checkm
    &&
    &&&
    &
    && \checkm &&
    &&&& \checkm &&&
    && \pie{2}\fp{} &&&&&&&
    & \pie{3} 
    & \pie{4} 
    \\

    DangSan~\cite{kouwe2017dangsan}
    & '17 &
    &&&
    &&&
    & \checkm
    &&
    &&&
    &
    && \checkm &&
    &&& \checkm & \checkm &&&
    && \pie{2}\fp{} &&&&&&& 
    & \piev{3} 
    & \pie{4} 
    \\

    MSan~\cite{stepanov2015memorysanitizer}
    & '15 & \checkm
    &&&
    &&&
    &
    && \checkm
    &&&
    &
    && \checkm &&&& \checkm &&&&&
    &&& \pie{3}\fp{} &&&&&& 
    & \piev{3} 
    & \pie{4} 
    \\

    CaVer~\cite{lee2015caver}
    & '15 &
    &&&
    &&&
    &
    &&
    & \checkm &&
    &
    & \checkm & \checkm &&&&&& \checkm &&& \checkm
    &&&& \pie{2} &&&&&
    & \pie{3}
    & \pie{4}
    \\

    TypeSan~\cite{haller2016typesan}
    & '16 &
    &&&
    &&&
    &
    &&
    & \checkm &&
    &
    & \checkm & \checkm &&&&& \checkm &&&& \checkm
    &&&& \pie{2} &&&&&
    & \piev{3}
    & \pie{4} 
    \\

    HexType~\cite{jeon2017hextype}
    & '17 &
    &&&
    &&&
    &
    &&
    & \checkm &&
    &
    & \checkm & \checkm &&
    &&&& \checkm &&& \checkm
    &&&& \pie{3} &&&&&
    & \piev{4}
    & \pie{4}
    \\

    Loginov et al.~\cite{loginov2001debugging}
    & '01 &
    &&&
    &&&
    &
    &&
    && \checkm &
    &
    & \checkm &&& 
    && \checkm &&&&& 
    &&&&& \pie{2}\fp{} &&&&
    & \pie{0} 
    & \na
    \\

    LLVM TySan~\cite{tysan}
    & '17 & \checkm
    &&&
    &&&
    &
    &&
    && \checkm &
    &
    & \checkm & \checkm &&&& \checkm &&&&&
    &&&&& \pie{3} &&&&
    & \na
    & \pie{2}
    \\

    Clang CFI~\cite{clang-cfi}
    & '15 & \checkm
    &&&
    &&&
    &
    &&
    & \checkm & \checkm &
    &
    & \checkm & \checkm &&
    & \checkm &&&&&& \checkm
    &&&& \pie{1} && \pie{4} &&&
    & \piev{4}
    & \pie{4}
    \\

    HexVASAN~\cite{biswas2017vasan}
    & '17 &
    &&&
    &&&
    &
    &&
    &&& \checkm
    &
    && \checkm &&&&&& \checkm &&& \checkm
    &&&&&&& \pie{4}\fp{} &&
    & \piev{4} 
    & \pie{4}
    \\

    UBSan~\cite{ubsan}
    & '12 & \checkm
    &&&
    &&&
    &
    &&
    & \checkm & \checkm &
    & \checkm
    & \checkm &&&
    & \checkm &&&&&& \checkm
    &&&& \pie{1} && \pie{4} && \pie{4} & \pie{2}
    & \piev{3}
    & \pie{4}
    \\
    \hline
  \end{tabular}
  }
  }
  \vspace{1em}
  \caption[overview]{
    Overview of sanitizers
    \\
    \vspace{1em}
    \parbox{22em}{
      Spatial safety violation\\
      \indent\piel{1} \textnormal{No stack/global var. overflow detection}\\
      \indent\piel{2} \textnormal{No overflow to padding detection}\\ 
      \indent\piel{3} \textnormal{No intra-object overflow detection}\\
      \\
      Temporal safety violation\\
      \indent\piel{1} \textnormal{Dangling pointer identified at compile-time}\\
      \indent\piel{2} \textnormal{No protection for pointers to reused memory\\ or register-stored pointers}\\
      \indent\piel{3} \textnormal{No protection for pointers to stack variables\\ or integer-typed pointers}\\
      \\
      Use of uninitialized variables\\
      \indent\piel{2} \textnormal{Fair coverage}\\
      \indent\piel{3} \textnormal{Good coverage}
    }
    \parbox{22em}{
      \vspace{0em}
      Bad-casting\\
      \indent\piel{1} \textnormal{Polymorphic class support only}\\
      \indent\piel{2} \textnormal{Non-polymorphic class support}\\
      \indent\piel{3} \textnormal{Better but incomplete run-time type tracing}\\
      \\
      Load type mismatch\\
      \indent\piel{2} \textnormal{Scalar type granularity}\\
      \indent\piel{3} \textnormal{Incomplete run-time type tracing}\\
      \\
      Other undefined behavior\\
      \indent\piel{2} \textnormal{Partial coverage}\\
      \\
      \indent\celll{} \textnormal{Known false positives}\\
      \\
    }
    \parbox{12em}{
      Performance overhead\\
      \indent\piel{0} \textnormal{Over 10x}\\
      \indent\piel{2} \textnormal{Up to 10x}\\
      \indent\piel{3} \textnormal{Up to 3x}\\
      \indent\piel{4} \textnormal{Up to 10\%}\\
      \indent\pievl{4} \textnormal{Verified (see Appendix)} \\
      \\
      Memory overhead\\
      \indent\piel{0} \textnormal{Over 10x}\\
      \indent\piel{2} \textnormal{3x to 10x}\\
      \indent\piel{4} \textnormal{1x to 3x}\\
      \\
      \indent\hspace{-0.5pt}\textnormal{\na{}}~~\textnormal{Data not available}\\
      \\
    }
  \vspace{-2.0em}
  }
  \label{tab:overview}
\end{table*}

Table~\ref{tab:overview} summarizes the features of sanitizers that are either
being actively maintained (as open source projects or commercial products), or
that were published at academic conferences.  Some of the tools we included were
originally designed as an exploit mitigation, and therefore do not have a
built-in error reporting mechanism. However, these tools do fit our definition
of a sanitizer (cf. Section~\ref{sec:motivation:mitigations_vs_sanitizers}) as
they can pinpoint the exact location of the vulnerable code, and they can
provide useful feedback if used in conjunction with a debugger.  We excluded
Intel Inspector XE~\cite{intel-inspectorxe}, ParaSoft
Insure++~\cite{parasoft-insure}, Micro Focus DevPartner
Studio~\cite{microfocus-devpartner}, and UNICOM Global
PurifyPlus~\cite{purifyplus}, because the lack of public information about these
sanitizers does not permit an accurate comparison.

For every sanitizer, the table shows which bugs it finds, which techniques it
uses to find those bugs, and which metadata storage scheme (if any) it uses. The
pie marks represent our assessment of how effective the sanitizer is, and how
efficient it is in terms of run-time and memory overheads. A colored cell
indicates that the sanitizer is known to produce false positives. We discuss the
reasons for these false positives in Section~\ref{sec:eval:accuracy}.  We
verified the reported performance numbers for $10$ of these tools (i.e., those
that have their performance overhead cells in Table~\ref{tab:overview} marked
with an asterisk) by running them on the same experimental platform using the
same set of benchmarks. We report the exact performance numbers for these tools
in Appendix~\ref{sec:perf-validation}.

\subsection{False Positives} \label{sec:eval:accuracy}

The practicality of a sanitizer primarily depends on how accurately it reports
bugs. A developer that uses a sanitizer wants to minimize the time spent on
reviewing its bug reports. The most desirable property for a sanitizer is
therefore that it reports no false positives (i.e., all bugs it reports are
truly bugs), while false negatives (i.e., the sanitizer finds all possible bugs)
are a secondary concern.  We identified the following recurring problems that
can lead to false positive detections.

The most frequently recurring problem is that sanitizers often implement a bug
finding policy or mechanism which is stricter than either the language standard
or the de facto standard.  The de facto standard includes widely-followed
programming practices that do not necessarily comply with the language standard,
even though they result in bug-free code~\cite{memarian2016defacto}. One could
therefore argue that reporting behavior that does not comply with the de facto
standard as a bug constitutes a false positive detection.

Older per-object bounds trackers such as J\&K disallow the creation of OOB
pointers that point beyond the end of an array~\cite{jones1997bounds}. This
design decision is compatible with the language standard, but not with the de
facto standard. Creating OOB pointers is common in real-world programs, and does
not lead to problems unless the program dereferences the OOB pointer. Subsequent
per-object bounds trackers such as CRED allow programs to create OOB pointers.

Programs that store temporarily OOB pointers can also cause problems for
dangling pointer
checkers~\cite{kouwe2017dangsan,younan2015freesentry,lee2015dangnull}. These
tools do not recognize the intended referent of an OOB pointer and therefore
fail to register the pointer to the correct object. If the intended referent
gets deallocated, the dangling pointer checker will not invalidate the OOB
pointer.  Worse, if the temporarily OOB pointer happened to point to a valid
object (different from its intended referent) when it was registered, then the
pointer checker will erroneously invalidate the pointer when the program
deallocates that different object.

Pointers that temporarily go OOB can also cause problems in low-fat
pointer-based bounds checkers~\cite{duck2016lowfatheap, duck2017lowfatstack},
which perform bounds checks whenever a pointer escapes from a function. If an
OOB pointer is passed to a function that converts that pointer back to its
original in-bounds value, the checker will have raised a false positive warning.

Uninitialized memory read detectors raise warnings when the program reads
uninitialized memory. The language standard allows this, as long as the
uninitialized values are not used in the program.

Some pointer type error checkers fail to capture the effective type of an object
under certain circumstances. For example, if a memory region is repurposed using
placement new in \C++, these checkers may fail to update or invalidate the type
metadata associated with that region. This can lead to false positive detections
when the old type is used for type checking.

Type checkers such as Loginov et al.'s pointer use
monitor~\cite{loginov2001debugging} and HexVASAN~\cite{biswas2017vasan} require
that the source and destination types are identical. The language standard's
aliasing rule defines a set of permitted conversions between non-identical
types, however.

\subsection{False Negatives} \label{sec:analysis:falsenegatives}

False negatives (i.e., failing to report bugs that are in scope) happen due
to discrepancies between the bug finding policy and the mechanism that
implements the policy. We identified several bug finding mechanisms that do not
fully cover all of the bugs that are supposed to be covered by the policy.

Spatial safety violation detection mechanisms based on red-zone insertion and
guard pages only detect illegal accesses to the red-zone or guard page directly
adjacent to the intended referent of that access. Memory accesses that target a
valid object beyond the red-zone or guard page are not detected.  These same
mechanisms also fail to detect intra-object overflows because they do not insert
red-zones or guard pages between subobjects in the same parent object.

Spatial safety violation detectors that use tagged pointers may round up the
allocation sizes for newly allocated objects to the nearest power of
two~\cite{akritidis2009baggy}, or to the nearest supported allocation
size~\cite{duck2016lowfatheap, duck2017lowfatstack}. The bounds checks performed
by these detectors enforce allocation bounds, rather than object bounds. Thus,
if an illegal memory access targets the padding added to an object, it will not
be reported.

Per-object bounds tracking tools do not detect intra-object overflows because
they cannot (always) distinguish object pointers from subobject pointers. This
happens, e.g., when a parent object has a subobject as its first member. This
subobject is located at the same memory address as the parent object.

Temporal safety violation detection mechanisms based on location-based checking
or guard pages cannot detect dereferencing of dangling pointers whose target has
been reused for a new memory allocation. This problem can be mitigated by
delaying memory reuse for a limited time, or eliminated if a pool allocation
analysis can determine when deallocated is no longer
accessible~\cite{dhurjati2006dangling}. Pool allocation analysis is only
available at compile time, when sufficient type information is available,
however.

Guard page-based temporal safety violation detectors cannot invalidate local
variables that have gone out of scope. These local variables are stored in stack
frames. These frames cannot be replaced by guard pages because they usually
share memory pages with other frames that are still in use. Consequently, guard
page-based techniques cannot detect use-after-scope and use-after-return
vulnerabilities.

Temporal safety violation detectors based on dangling pointer invalidation only
invalidate pointers that are stored in memory. Dangling pointers stored in
registers are not invalidated, even if the program eventually copies them into
memory.

Most uninitialized memory use detectors approximate the language standard by
considering a value as ``used'' only in limited circumstances, such as when it
is passed to a system call, or when it is used in a branch condition.

Pointer type error detectors such as TySan~\cite{tysan} also conservatively
approximate the effective type rules of the language standard, thus failing to
detect bugs involving objects of a type unknown to their system.

Some sanitizers fail to recognize pointers that are cast to integers or copied
via \code{memcpy}. Identity-based access checkers that use per-pointer metadata,
for example, regularly fail to propagate pointer bounds across these constructs.
This problem also affects sanitizers that tag dangling pointers by instrumenting
stores of pointer-typed variables, but miss pointers temporarily cast to
integers or copied in a type-unsafe manner.

\subsection{Incomplete Instrumentation}

Sanitizers that instrument programs statically cannot fully support programs
that generate code at run time (e.g., just-in-time compilers) or programs that
interact with external libraries that cannot be instrumented (e.g., because
their source code is not available). Some sanitizers that instrument programs at
the compiler IR level also do not support programs containing inline assembly
code because the compiler front-end does not translate such code into compiler
IR code.  In all of these cases, the sanitizer might fail to insert checks,
potentially leading to false negatives. For example, if a program accesses
memory from within a block of dynamically generated code, a spatial safety
violation sanitizer will generally not be able to verify whether the memory
access is legal.

Moreover, the sanitizer might also fail to emit the necessary instructions to
update metadata. This is particularly problematic for sanitizers that need
to propagate metadata explicitly (e.g., disjoint per-pointer metadata). For
example, if a program copies a pointer with disjoint metadata to a new memory
location from within an external non-instrumented library, then the sanitizer
will not copy the metadata for the source pointer. Without proper
metadata propagation, the sanitizer might generate false negatives (because
metadata might be missing from the store) or false positives (because the
metadata might be outdated).

These problems can be overcome by embedding the sanitizer at run time instead,
using a dynamic binary instrumentation framework. These frameworks cannot
provide accurate type information, however, and consequently do not support
certain types of sanitizers (e.g., pointer casting monitors).

\subsection{Thread Safety}

Sanitizers that maintain metadata for pointers and objects can incur both false
positives and false negatives in multi-threaded programs. This can happen
because they might access metadata structures in a thread-unsafe way, or because
the sanitizer does not guarantee that it updates the metadata in the same
transaction as program's atomic updates to its associated pointers or
objects. The former problem affects FreeSentry~\cite{younan2015freesentry} and
makes the sanitizer unable to support multi-threaded programs. The latter
problem affects Intel Pointer Checker~\cite{intel-pointerchecker}, and Intel
MPX~\cite{intel-mpx} among others. These sanitizers allow pointers or objects to
go out of sync with their metadata if the program concurrently updates and
accesses them outside of critical sections. Some sanitizers such as
Memcheck~\cite{seward2005memcheck} sidestep this issue by serializing the
execution of multi-threaded programs, thereby always atomically updating
metadata along with pointers and objects associated with it.

\subsection{Performance Overhead} \label{sec:cost}  \label{sec:cost:perf}

The run-time performance requirements for sanitizers are not as stringent
as those for exploit mitigations. While the latter typically only see real-world
deployment if their run-time overhead stays below 5\%~\cite{szekeres.etal+13},
we observed that sanitizers incurring less than 3x overhead are widely used in
practice. In some cases, such as when the source code for a program is not
(fully) available, or if the program generates code on-the-fly, even larger
overheads of up to 20x are acceptable. Yet, there are good reasons to try to
minimize a sanitizer's overhead. One reason in particular is that the faster a
sanitizer becomes, the faster a sanitization-enabled program can be fuzzed. This
in turn allows the fuzzer to explore more code paths before it stops making
meaningful progress (cf. Section~\ref{sec:sanitizers:drivers}).

The primary contributors to a sanitizer's run-time overhead are its checking,
metadata storage and propagation, and run-time instrumentation cost. The
run-time instrumentation cost is zero for most sanitizers, because they
instrument programs statically (at compile time). For sanitizers that use
dynamic binary instrumentation, however, the run-time instrumentation cost can
be very high.  Valgrind's Memcheck~\cite{seward2005memcheck}, for example, incurs 25.7x
overhead on the SPEC2000 benchmarks. 4.9x of this run-time overhead can be
attributed to Valgrind itself~\cite{nethercote2007valgrind}, the DBT framework
Memcheck is based on.

The metadata storage and propagation cost primarily depends on the metadata
storage scheme. In general, embedded metadata and tagged or fat pointers are the
most efficient storage schemes because they cause less cache pressure than
disjoint/shadow metadata storage schemes. Embedded metadata and tagged/fat
pointers have the additional advantage that their metadata automatically
propagates when an object or pointer is copied. Using tagged or fat pointers is
problematic in programs that cannot be fully instrumented, however
(cf. Section~\ref{sec:metadata:pointer}). The one exception is low-fat
pointer-based bounds tracking~\cite{duck2016lowfatheap, duck2017lowfatstack},
where the metadata is stored implicitly in the tagged pointer so that the tagged
pointer can still be dereferenced in non-instrumented libraries. In practice, we
observe that disjoint/shadow metadata storage schemes are preferred over tagged
and fat pointers, despite the fact that they cause more cache pressure and
require explicit metadata propagation when objects or pointers are copied.

The checking cost is strongly correlated with the sanitizer's checking
frequency, which, in turn, strongly depends on the type of sanitizer.
Since memory error detectors generally require coverage of all memory accesses
or pointer arithmetic operations performed by a program, they introduce more
overhead than other tools such as type casting checkers that monitor a smaller
set of operations.  Some memory error detection tools provide selective
instrumentation, e.g., to monitor memory writes only, to achieve better
performance at the cost of reduced coverage.

\subsection{Memory Overhead} \label{sec:cost:mem}

Sanitizers that increase the allocation sizes for memory objects, or that use
disjoint or shadow metadata storage schemes have sizable memory footprints. This
can be problematic on 32-bit platforms, where the amount of the addressable
memory space is limited.  ASan~\cite{serebryany2012addresssanitizer}, for
example, inserts red zones into every memory object and maintains a
direct-mapped shadow bitmap to store addressability information. Consequently,
ASan increases the memory usage of the SPEC2006 benchmarks by 3.37x on
average. Guard page-based memory safety sanitizers, such as Electric
Fence~\cite{perens1993electric} and PageHeap~\cite{microsoft2000pageheap},
insert entire memory pages at the end of dynamically allocated objects, and
therefore have even bigger memory footprints.  In general, however, most
sanitizers increase the program's memory footprint by less than 3x on average,
even if the sanitizer stores metadata for every object or pointer in the
program.

\section{Deployment}

We studied the current use of sanitizers. Our goals were to determine (i)~what
sanitizers are favored by practitioners and (ii)~how they differ from those that
are not.

\subsection{Methodology}

\paragraph*{\textbf{Popular GitHub repositories}}
We compiled a list of the top 100 C and top 100 \C++ projects on GitHub and
examined their build and test scripts, GitHub issues, and commit histories. Most
of the sanitizers we reviewed need to be integrated into the tested program at
compile time. A program's build configuration would therefore reveal whether it
is regularly sanitized. Our examination of the test suites and testing scripts
further showed which sanitizers can be enabled during testing.  Contrary to the
build system/configuration files, references to sanitizers that instrument
programs at run time (e.g., Memcheck) would show up here.

\paragraph*{\textbf{Sanitizer web pages}}
We examined the web sites for sanitizer tools and looked for explicit
references to projects using the sanitizer and acknowledgments of bug
discoveries.

\paragraph*{\textbf{Search trends}}
We examined and compared search trends for different sanitizers. We used
AddressSanitizer as the baseline in the search trends as our study indicates
that it is currently the most widely deployed sanitizer.

\subsection{Findings}

\paragraph*{\textbf{AddressSanitizer is the most widely adopted sanitizer}}
We found that AddressSanitizer (ASan) is used in~24 and~19 of the most
popular C and~\C++ projects on GitHub respectively. We believe that this
popularity can be attributed to several of ASan's strengths: (i)~ASan detects
the class of bugs with the highest chance of exploitation (memory safety
violations), (ii)~ASan is highly compatible because it does not incur additional
false positives when the program is not fully instrumented (e.g., because the
program loads non-instrumented shared libraries), (iii)~ASan has a low false
positive rate in general and false positives that do occur can be suppressed by
adding annotations to the program, or by adding the location where the false
positive detection occurs to a blacklist, (iv)~ASan is integrated into
mainstream compilers. Enabling ASan therefore requires only trivial changes to
the tested program's build system, and (v)~ASan scales to large programs such as
the Chromium and Firefox web browsers. A weakness of ASan, and of other
sanitizers that combine location-based checking with red-zone insertion, is that
it produces false negatives.

One interesting observation is that DBT-based memory safety sanitizers such as
Memcheck and Dr. Memory have nearly identical strengths. Additionally, these
sanitizers can always instrument the full program even if part of the program's
source code is not available. Yet, our study shows that while Memcheck was
popular before ASan was introduced into LLVM and GCC, its real-world use now
trails that of ASan. Dr. Memory, being a much more recent tool, never achieved
the same level of adoption than either of the competitors.

\paragraph*{\textbf{The adoption rate for other LLVM-based sanitizers is lower}}
MSan and UBSan have also seen adoption, mainly due to increased attention
towards vulnerabilities such as uninitialized memory use and integer overflows.
However, users frequently report high false positive rates and avoiding them
requires significant effort. In fact, developers have to go to great lengths to
apply those sanitizers to large projects like the Chromium web browser.  To
avoid false positives for MSan, the entire program must be instrumented. In
Chromium's case, this means that MSan must be inserted into the web browser
itself, as well as in all of its dependencies. For UBSan, the developers
maintain a long list of suppressions that most notably suppresses all detections
in the entire V8~JavaScript engine.

\subsection{Deployment Directions}
The deployment landscape hints at the desirable properties of a sanitizer.
First, all the deployed sanitizers are easy to use. Specifically, they can be
enabled via a compiler flag (Clang sanitizers) or can be applied to any binary
(Memcheck). Second, the false positive rate and adoption are inversely related,
i.e., fewer false positives means higher adoption (ASan and Memcheck vs. MSan
and UBSan). Third, performance overhead is not a primary concern (Memcheck is
used), but is avoided when a faster alternative is available (Memcheck vs.
ASan).

Our own experience of applying sanitizers to the SPEC benchmarks shows that
research prototypes suffer even more from false positives than widely deployed
sanitizers. ASan successfully ran all the benchmarks, correctly reporting known
bugs in SPEC. Memcheck ran all benchmarks except for \code{447.dealII}, which
takes more than 48 hours to finish. In contrast, SoftBound+CETS fails to run
many of the benchmarks raising false alarms, due to strictness (e.g., not
supporting integer-pointer casts) and compatibility (e.g., failure to update
bounds for pointers created in uninstrumented libraries) issues.  LFP failed to
run several benchmarks, because the assumed invariant that OOB pointers do not
escape the creating function is too strict. DangSan provides their own patches
to circumvent incorrect invalidation of pointers.

Developers who want to sanitize memory safety issues in their projects can pick
up ASan or Memcheck without much effort. However, they should be aware that
these tools do not detect all classes of memory safety violations. Developers
who want to adopt MSan and UBSan should expect continued efforts
for recompilation of all the dependencies and/or for blacklisting and annotation
to weed out false positives. For the vulnerabilities not covered by these
popular sanitizers (e.g., intra-object overflow and type errors caused by type
punning), developers currently have no viable option. Further research in
this area is required, because existing research prototypes do not scale to
real-world code bases.

\section{Future Research and Development Directions}
\label{sec:future}

\subsection{Type Error Detection}

Most of the publicly available sanitizers that detect pointer type errors do so
by monitoring casting operations. Such tools can detect illegal casts, but due
to type-punning constructs such as unions and \code{reinterpret\_cast} casts, it
is still possible for pointers to have an illegal type when the program
dereferences them. Pointer use monitoring can solve this problem because it
tracks the effective types of every storage location, and can therefore detect
illegal dereferences of pointers that were derived through type
punning. Unfortunately, only one publicly available tool monitors pointer uses
(LLVM TySan) and it is still a research prototype~\cite{tysan}. We were not able
to get TySan to work on non-trivial programs. TySan also has high memory
overhead because of its large metadata. Pointer use monitoring therefore remains
an interesting area of research.  Advances in pointer use monitoring will help
address the general problem of accurate pointer type error detection, which
other sanitizers face as well. For example, variadic function misuse sanitizers
will benefit from this, allowing precise detection of mismatches in variadic
argument types.

\subsection{Improving Compatibility}

Although there exist other memory vulnerability sanitizers with better
precision, AddressSanitizer (ASan) is by far the most deployed sanitizer. We
believe that the primary reason for ASan's wider deployment is its excellent
compatibility with the de facto language standards, and with partially
instrumented programs. We encourage future research and development efforts to
make other sanitizers equally compatible.

\subsection{Composing Sanitizers}

Sanitizers typically detect one particular class of security bugs.  Embedding
multiple sanitizers is unfortunately not possible at present because existing
sanitizers use a variety of incompatible metadata storage schemes (several of
which change the pointer representation). Current practice is therefore to test
the program multiple times, once with each sanitizer. This requires developers
to make an additional investment in time and effort.

We encourage further research into efficient metadata storage schemes that are
sufficiently generic to support a wide variety of
sanitizers~\cite{haller2016metalloc,kroes2017midfat}, and into sanitizers that
build on such metadata storage schemes. This problem could also be addressed by
using multi-variant execution systems to run multiple variants of the same
program in parallel on the same inputs. Different sanitizers can be embedded in
each variant, allowing incompatible sanitizers to run in
parallel~\cite{xu2017bunshin,pina2018freeda}.

\subsection{Hardware Support}

Hardware features can lower the run-time performance impact of sanitization,
improve bug detection precision, and alleviate certain compatibility issues. The
idea of using special hardware instructions to accelerate memory safety
violation detection has already been extensively
explored~\cite{devietti2008hardbound, nagarakatte2012watchdog,
  nagarakatte2014watchdoglite}. Recent Intel CPUs even include an ISA extension
called memory protection extension (MPX) built for memory error
detection~\cite{intel-mpx}. Intel MPX improves the speed of the software
implementation of the same mechanism, though there is still room for
improvement~\cite{oleksenko2017mpxexplained}.

In addition, hardware features could improve compatibility and precision. For
example, ARM's virtual address tagging allows top eight bits of the virtual
address be ignored by the processor. This can be used to implement the tagged
pointer scheme which does not incur binary compatibility issues, because
dereferencing a tagged pointer in an non-instrumented library no longer leads to
processor faults. This tag also propagates back to the instrumented library,
potentially increasing the bug detection precision. Hardware-assisted
AddressSanitizer, being developed as of writing, uses this feature to detect
both spatial and temporal memory safety violations at lower performance and
memory costs than AddressSanitizer~\cite{hwasan}.

\subsection{Kernel and Bare-Metal Support}

Sanitizers have traditionally only been available for user-space applications.
Lower-level software such as kernels, device drivers and hypervisors is
therefore missing out on the security benefits of sanitization. Unfortunately,
security bugs may have the most disastrous consequences in such low-level
software. Efforts are ongoing to remedy this problem. Projects led by Google,
for example, are bringing AddressSanitizer and MemorySanitizer to the Linux
kernel~\cite{kasan, kmsan}. We encourage these efforts and hope to see other
classes of sanitizers adopted to lower level software. One challenge for this is
to reduce the sanitizer's memory footprint. While memory overheads of 3x or more
are acceptable in user-space applications for 64-bit platforms, such overheads
could be a problem for lower level software on 32-bit architectures. The Linux
kernel, in particular, is often compiled and run on 32-bit platforms (e.g., on
IoT devices).

{\balance
\bibliographystyle{unsrt}
\bibliography{references}
}

\begin{appendices}

\section{}
\label{sec:perf-validation}

We measured the run-time performance overhead of $10$ tools by running all $19$
SPEC CPU2006 C/\C++ benchmarks (or all $7$ \C++ benchmarks for type casting
sanitizers) on a single experimental platform. To assist future sanitizer
developers in measuring the relative overhead of their tool, we open-sourced our
fully automated build and benchmarking scripts at
https://placeholder.edu/spec-scripts\footnote{The scripts will be made available
  upon acceptance of this paper}.

\subsection{Scope}

We included sanitizers that are either actively maintained and/or were published
within a decade. For sanitizers that are not publicly available, we sent the
authors a request for source code access. However, the authors either did not
respond~\cite{lee2015dangnull, caballero2012undangle, younan2010paricheck}, or
refused our request because of licensing
restrictions~\cite{intel-pointerchecker, akritidis2009baggy} or code quality
concerns~\cite{dang2017oscar, younan2015freesentry}. We excluded several
sanitizers that either failed to compile or run more than half of the
benchmarks~\cite{nagarakatte2009softbound, nagarakatte2010cets, tysan}, or do
not support our experimental platform~\cite{microsoft2000pageheap}. CaVer caused
several instrumented binaries to run significantly faster than the baseline
binaries~\cite{lee2015caver}. Since these speedups were not reported in the
original paper, and since we did not have time to properly investigate the
cause of these speedups, we decided to exclude CaVer from our evaluation. In the
end, we evaluated 10 tools.

\subsection{Experimental Setup}

We conducted all experiments on a host equipped with an Intel Xeon E5-2660~CPU
with 20MB cache and 64GB RAM running 64-bit Ubuntu~14.04.5 LTS.
Unless stated otherwise, we used the system default libraries installed in the
OS distribution.

\subsection{Results}

We ran each benchmark three times with the sanitizer and report the median for
each run normalized to the baseline. We used the median of three runs without
the sanitizer as the baseline result. We also report and give details about
false positives encountered while running benchmarks in this section.
Table~\ref{tab:perf} summarizes the overheads and false positives.

\newcolumntype{x}{>{\centering\arraybackslash}p{2em}}

\newcommand\diagonal[2]{
  \multicolumn{1}{p{#1}#2}{\hskip-\tabcolsep
  $\vcenter{\begin{tikzpicture}[baseline=0,anchor=south west,inner sep=.1em]
  \path[use as bounding box] (.01em,0) rectangle (#1+2*\tabcolsep,\baselineskip);
  \node[minimum width={#1+2*\tabcolsep-.04em},minimum height=\baselineskip+\extrarowheight-.06em] (box) {};
  \draw[light-gray] (box.south west) -- (box.north east);
  \end{tikzpicture}}$\hskip-\tabcolsep}
}
\newcommand\notapplicable{\diagonal{2em}{?}}
\newcommand\notapplicabler{\diagonal{2em}{|}}

\begin{table*}[!ht]
  \setlength{\tabcolsep}{0.2em}
  \setlength{\extrarowheight}{0.2em}
  \centering
  \caption{Normalized Performance Overheads and False Positives}
  \label{tab:perf}
  {\rowcolors{4}{lightgray!20}{white}
  \begin{tabular}{| l | x?x?x?x?x?x?x?x?x?x?x?x | x?x?x?x?x?x?x | x | }
    \hline
    & \multicolumn{12}{c|}{\thead{SPEC CPU2006 INT}}
    & \multicolumn{7}{c|}{\thead{SPEC CPU2006 FP}}
    & \\
    & \rotatebox{90}{400.perlbench}
    & \rotatebox{90}{401.bzip2}
    & \rotatebox{90}{403.gcc}
    & \rotatebox{90}{429.mcf}
    & \rotatebox{90}{445.gobmk}
    & \rotatebox{90}{456.hmmer}
    & \rotatebox{90}{458.sjeng}
    & \rotatebox{90}{462.libquantum}
    & \rotatebox{90}{464.h264ref}
    & \rotatebox{90}{471.omnetpp}
    & \rotatebox{90}{473.astar}
    & \rotatebox{90}{483.xalancbmk}
    & \rotatebox{90}{433.milc}
    & \rotatebox{90}{444.namd}
    & \rotatebox{90}{447.dealII}
    & \rotatebox{90}{450.soplex}
    & \rotatebox{90}{453.povray}
    & \rotatebox{90}{470.lbm}
    & \rotatebox{90}{482.sphinx3}
    & \rotatebox{90}{Geometric Mean}\\
    \hline
    & C & C & C & C & C & C & C & C & C & \C++ & \C++ & \C++ & C & \C++ & \C++ & \C++ & \C++ & C & C & \\
    \hline
    Memcheck~\cite{seward2005memcheck}
    & 39.7 & 14.1 & 21.0 & 4.34 & 24.9 & 23.9 & 26.8 & 9.81 & 31.9 & 17.5 & 11.0 & 28.5 & 13.4 & 24.4 & \notapplicable & 10.9 & 47.8 & 23.0 & 34.0 & 19.6
    \\
    ASan~\cite{serebryany2012addresssanitizer}
    & 4.27 & 1.71 & 2.32 & 1.47 & 2.17 & 2.01 & 2.31 & 1.39 & 2.37 & 2.24 & 1.62 & 2.77 & 1.44 & 1.58 & 2.48 & 1.65 & 2.99 & 1.07 & 1.93 & 1.99
    \\
    Low-fat Pointer~\cite{duck2016lowfatheap, duck2017lowfatstack}
    & 2.52\fp{} & 1.76 & 2.27\fp{} & 1.40 & 1.87 & 2.49 & 1.73 & 1.92 & 2.64\fp{} & 1.46 & 1.76 & 1.96
    & 1.54 & 1.95\fp{} & 2.16 & 1.94\fp{} & 2.44 & 1.47 & 2.06
    & 1.93
    \\
    DangSan~\cite{kouwe2017dangsan}
    & 3.63\fp{} & 1.02 & 1.55 & 1.50 & 1.13 & 1.01 & 1.03 & 0.92 & 0.99 & 6.83 & 1.51 & 2.28 & 1.34 & 1.00 & 1.25 & 1.09\fp{} & 1.56 & 1.00 & 1.01 & 1.40
    \\
    MSan~\cite{stepanov2015memorysanitizer}
    & 3.50 & 2.04 & 3.51\fp{} & 2.16 & 2.51 & 3.68 & 3.49 & 1.93 & 3.40 & 2.24 & 1.95 & 2.21 & 1.99 & 1.96 & 2.62 & 1.95\fp{} & 3.29 & 2.20 & 2.85 & 2.53
    \\
    TypeSan\cite{haller2016typesan}
    & \notapplicable & \notapplicable & \notapplicable & \notapplicable & \notapplicable & \notapplicable & \notapplicable & \notapplicable & \notapplicable & 1.64 & 0.99 & 1.41 & \notapplicable & 0.99 & 1.81 & 1.00 & 1.24 & \notapplicable & \notapplicabler & 1.26
    \\
    HexType~\cite{jeon2017hextype}
    & \notapplicable & \notapplicable & \notapplicable & \notapplicable & \notapplicable & \notapplicable & \notapplicable & \notapplicable & \notapplicable & 1.13 & 1.01 & 1.08 & \notapplicable & 1.00 & 1.18 & 1.02 & 1.01 & \notapplicable & \notapplicabler & 1.06
    \\
    Clang CFI~\cite{clang-cfi}
    & \notapplicable & \notapplicable & \notapplicable & \notapplicable & \notapplicable & \notapplicable & \notapplicable & \notapplicable & \notapplicable & 1.46 & 1.00 & 1.16 & \notapplicable & 1.00 & 1.09 & 0.99 & 1.03 & \notapplicable & \notapplicabler & 1.09
    \\
    HexVASAN~\cite{biswas2017vasan}
    & 1.03 & 1.01 & 1.01 & 1.00 & 1.01 & 1.00 & 0.99 & 1.03 & 1.00 & 1.04\fp{} & 1.00 & 0.99 & 1.00 & 1.00 & 1.00 & 1.00 & 1.00 & 1.00 & 1.00 & 1.01
    \\
    UBSan~\cite{ubsan}
    & 2.20 & 2.33 & 1.98 & 1.90 & 3.04 & 6.68 & 2.94 & 2.87 & 4.31 & 4.26 & 2.22 & 5.37 & 2.38 & 2.08 & 9.16 & 2.67 & 3.01 & 1.24 & 3.02 & 2.97
    \\
    \hline
  \end{tabular}
  }
  \vspace{-1em}
\end{table*}

\subsubsection{Memcheck}

We used the official version of LLVM/Clang 6.0.0 to compile the baseline
binaries. We measured Memcheck's overhead by running Valgrind 3.13.0 with the
\code{--tool=memcheck} option. We excluded \code{447.dealII} as it does not
finish within 48 hours.

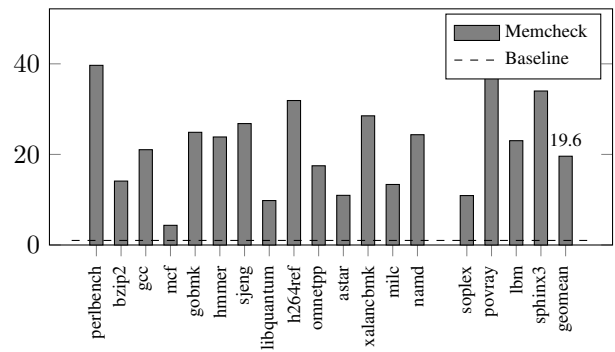
\pgfplotstableread[col sep=comma]{memcheck.csv}{\memcheck}
\pgfplotstablegetelem{0}{overhead}\of{\memcheck}
\pgfmathsetmacro{\gm}{\pgfplotsretval}

\begin{figure}[H]
  \centering
  \begin{tikzpicture}
    \begin{axis}[
      ybar, bar width=5pt, legend cell align=left,
      legend style={font=\scriptsize}, area legend,
      height=0.2\textheight, width=0.5\textwidth,
      symbolic x coords={left,perlbench,bzip2,gcc,mcf,gobmk,hmmer,sjeng,libquantum,h264ref,omnetpp,astar,xalancbmk,milc,namd,dealII,soplex,povray,lbm,sphinx3,geomean,right},
      x tick label style={font=\scriptsize,rotate=90},
      xtick pos=left,
      xtick=data,
      ]
      \addplot[fill=gray] table[x={benchmark}, y={overhead}] {\memcheck};
      \addplot[black,line legend,sharp plot,update limits=false,style=dashed]
      coordinates {(left, 1.0) (right, 1.0)};
      \node[black, above, style={font=\scriptsize}] at (axis cs: geomean,\gm) {\gm};
      \legend{Memcheck, Baseline}
    \end{axis}
  \end{tikzpicture}
\end{figure}

\subsubsection{AddressSanitizer}

We used the official version of LLVM/Clang 6.0.0 to compile the baseline
binaries, and compiled the AddressSanitizer binaries using the
\code{-fsanitize=address} flag for that same compiler. We patched several known
bugs in \code{400.perlbench} and \code{464.h264ref} to avoid crashing those
benchmarks early. We disabled detection of memory leaks and alloc/dealloc
mismatches.

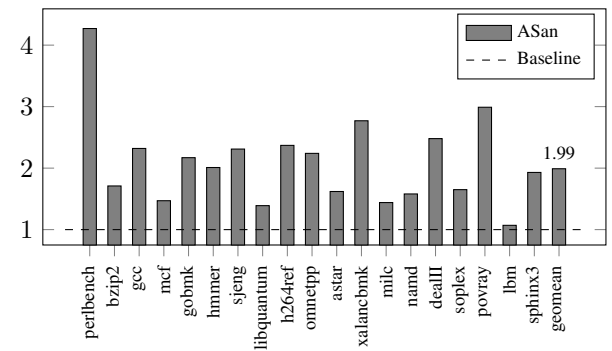
\pgfplotstableread[col sep=comma]{asan.csv}{\asan}
\pgfplotstablegetelem{0}{overhead}\of{\asan}
\pgfmathsetmacro{\gm}{\pgfplotsretval}

\begin{figure}[H]
  \centering
  \begin{tikzpicture}
    \begin{axis}[
      ybar, bar width=5pt, legend cell align=left,
      legend style={font=\scriptsize}, area legend,
      height=0.2\textheight, width=0.5\textwidth,
      symbolic x coords={left,perlbench,bzip2,gcc,mcf,gobmk,hmmer,sjeng,libquantum,h264ref,omnetpp,astar,xalancbmk,milc,namd,dealII,soplex,povray,lbm,sphinx3,geomean,right},
      x tick label style={font=\scriptsize,rotate=90},
      xtick pos=left,
      xtick=data,
      ]
      \addplot[fill=gray] table[x={benchmark}, y={overhead}] {\asan};
      \addplot[black,line legend,sharp plot,update limits=false,style=dashed]
      coordinates {(left, 1.0) (right, 1.0)};
      \node[black, above, style={font=\scriptsize}] at (axis cs: geomean,\gm) {\gm};
      \legend{ASan, Baseline}
    \end{axis}
  \end{tikzpicture}
\end{figure}

\subsubsection{Low-fat Pointer}

We used LFP's version of LLVM/Clang 4.0.0 to compile the baseline binaries and
used the \code{-fsanitize=lowfat} flag for that same compiler to generate the
LFP binaries. We used the default size encoding in the low-fat pointer
representation. As CPUs in our experimental platform do not have bit
manipulation extensions, we removed optimization flags using those extensions,
though they are enabled in author's build script. We disabled early program
termination after check failures to measure the overhead for benchmarks with
known false positives.

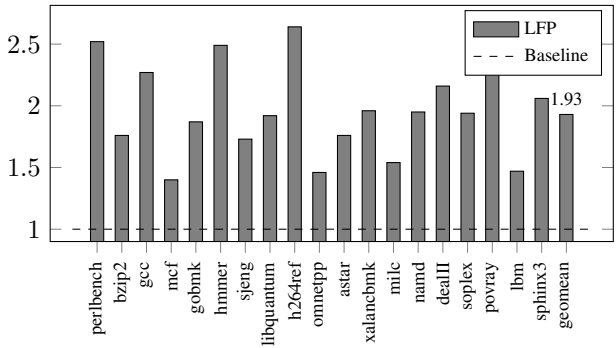
\pgfplotstableread[col sep=comma]{lowfat.csv}{\lowfat}
\pgfplotstablegetelem{0}{overhead}\of{\lowfat}
\pgfmathsetmacro{\gm}{\pgfplotsretval}

\begin{figure}[H]
  \centering
  \begin{tikzpicture}
    \begin{axis}[
      ybar, bar width=5pt, legend cell align=left,
      legend style={font=\scriptsize}, area legend,
      height=0.2\textheight, width=0.5\textwidth,
      symbolic x coords={left,perlbench,bzip2,gcc,mcf,gobmk,hmmer,sjeng,libquantum,h264ref,omnetpp,astar,xalancbmk,milc,namd,dealII,soplex,povray,lbm,sphinx3,geomean,right},
      x tick label style={font=\scriptsize,rotate=90},
      xtick pos=left,
      xtick=data,
      ymin=0.9,
      ]
      \addplot[fill=gray] table[x={benchmark}, y={overhead}] {\lowfat};
      \addplot[black,line legend,sharp plot,update limits=false,style=dashed]
      coordinates {(left, 1.0) (right, 1.0)};
      \node[black, above, style={font=\scriptsize}] at (axis cs: geomean,\gm) {\gm};
      \legend{LFP, Baseline}
    \end{axis}
  \end{tikzpicture}
\end{figure}

\subsubsection{DangSan}

We used the official version of LLVM/Clang 3.8.0 to compile the baseline
binaries, and the DangSan plugin for that same compiler to generate the DangSan
binaries.  DangSan requires linking with the GNU gold linker and link time
optimization (\code{-flto}) enabled, so we used gold and \code{-flto} to link
the baseline binaries as well. Similarly, since DangSan uses \code{tcmalloc} as
the default memory allocator, we enabled \code{tcmalloc} for the baseline
binaries too. We did not enable \code{-fsanitize=safe-stack} for the baseline
binaries, since it incurs overhead. To avoid false positives in
\code{450.soplex}, we applied the pointer unmasking patch provided by the
authors. We marked \code{400.perlbench} as having false positives based on their
metadata invalidation workaround present in \code{tcmalloc}.

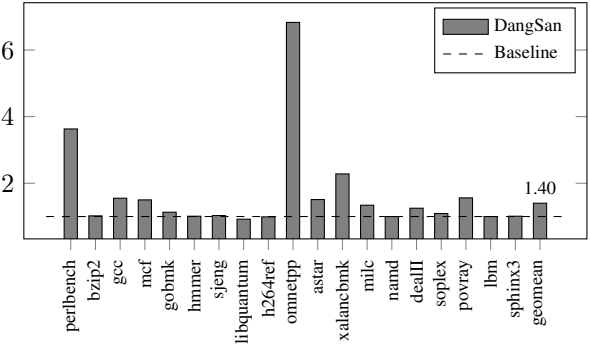
\pgfplotstableread[col sep=comma]{dangsan.csv}{\dangsan}
\pgfplotstablegetelem{0}{overhead}\of{\dangsan}
\pgfmathsetmacro{\gm}{\pgfplotsretval}

\begin{figure}[H]
  \centering
  \begin{tikzpicture}
    \begin{axis}[
      ybar, bar width=5pt, legend cell align=left,
      legend style={font=\scriptsize}, area legend,
      height=0.2\textheight, width=0.5\textwidth,
      symbolic x coords={left,perlbench,bzip2,gcc,mcf,gobmk,hmmer,sjeng,libquantum,h264ref,omnetpp,astar,xalancbmk,milc,namd,dealII,soplex,povray,lbm,sphinx3,geomean,right},
      x tick label style={font=\scriptsize,rotate=90},
      xtick pos=left,
      xtick=data,
      ]
      \addplot[fill=gray] table[x={benchmark}, y={overhead}] {\dangsan};
      \addplot[black,line legend,sharp plot,update limits=false,style=dashed]
      coordinates {(left, 1.0) (right, 1.0)};
      \node[black, above, style={font=\scriptsize}] at (axis cs: geomean,\gm) {\gm};
      \legend{DangSan, Baseline}
    \end{axis}
  \end{tikzpicture}
\end{figure}

\subsubsection{MemorySanitizer}

We used the official version of LLVM/Clang 6.0.0 to compile the baseline
binaries, and used the \code{-fsanitize=memory} flag for that same compiler to
generate the MemorySanitizer binaries. We used instrumented versions of
\code{libcxx} and \code{libcxxabi} when running \C++ benchmarks. This addresses
a false positive detection in \code{450.soplex}. We disabled early program
termination after check failures to measure the run-time overhead for
\code{403.gcc}.

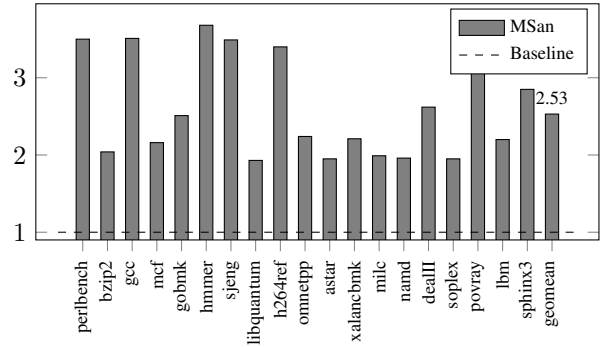
\pgfplotstableread[col sep=comma]{msan.csv}{\msan}
\pgfplotstablegetelem{0}{overhead}\of{\msan}
\pgfmathsetmacro{\gm}{\pgfplotsretval}

\begin{figure}[H]
  \centering
  \begin{tikzpicture}
    \begin{axis}[
      ybar, bar width=5pt, legend cell align=left,
      legend style={font=\scriptsize}, area legend,
      height=0.2\textheight, width=0.5\textwidth,
      symbolic x coords={left,perlbench,bzip2,gcc,mcf,gobmk,hmmer,sjeng,libquantum,h264ref,omnetpp,astar,xalancbmk,milc,namd,dealII,soplex,povray,lbm,sphinx3,geomean,right},
      x tick label style={font=\scriptsize,rotate=90},
      xtick pos=left,
      xtick=data,
      ymin=0.9,
      ]
      \addplot[fill=gray] table[x={benchmark}, y={overhead}] {\msan};
      \addplot[black,line legend,sharp plot,update limits=false,style=dashed]
      coordinates {(left, 1.0) (right, 1.0)};
      \node[black, above, style={font=\scriptsize}] at (axis cs: geomean,\gm) {\gm};
      \legend{MSan, Baseline}
    \end{axis}
  \end{tikzpicture}
\end{figure}

\subsubsection{TypeSan}

We used TypeSan's version of LLVM/Clang 3.9.0 to compile the baseline binaries
and we generated the TypeSan binaries using the \code{-fsanitize=typesan} flag
for that same compiler. TypeSan uses \code{tcmalloc} as its default memory
allocator, so we used an unmodified version of \code{tcmalloc} in the baseline
binaries too.

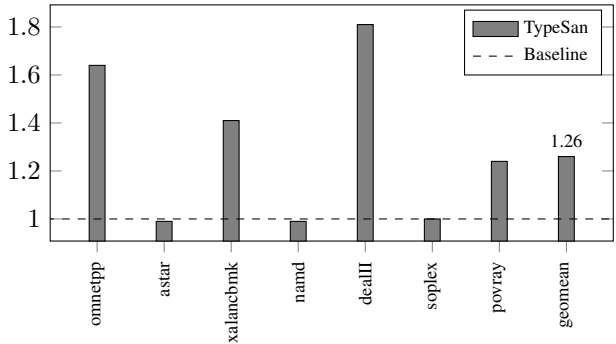
\pgfplotstableread[col sep=comma]{typesan.csv}{\typesan}
\pgfplotstablegetelem{0}{overhead}\of{\typesan}
\pgfmathsetmacro{\gm}{\pgfplotsretval}

\begin{figure}[H]
  \centering
  \begin{tikzpicture}
    \begin{axis}[
      ybar, bar width=6pt, legend cell align=left,
      legend style={font=\scriptsize}, area legend,
      height=0.2\textheight, width=0.5\textwidth,
      symbolic x coords={left,omnetpp,astar,xalancbmk,namd,dealII,soplex,povray,geomean,right},
      x tick label style={font=\scriptsize,rotate=90},
      xtick pos=left,
      xtick=data,
      ]
      \addplot[fill=gray] table[x={benchmark}, y={overhead}] {\typesan};
      \addplot[black,line legend,sharp plot,update limits=false,style=dashed]
      coordinates {(left, 1.0) (right, 1.0)};
      \node[black, above, style={font=\scriptsize}] at (axis cs: geomean,\gm) {\gm};
      \legend{TypeSan, Baseline}
    \end{axis}
  \end{tikzpicture}
\end{figure}

\subsubsection{HexType}

We applied HexType's patches to the official version of LLVM/Clang 3.9.0 and
used that compiler to generate both the baseline and the HexType binaries.  We
used the \code{-fsanitize=hextype} for the HexType binaries. We enabled all type
casting coverage and optimization features supported by HexType. This is
consistent with the experiments performed by the authors.

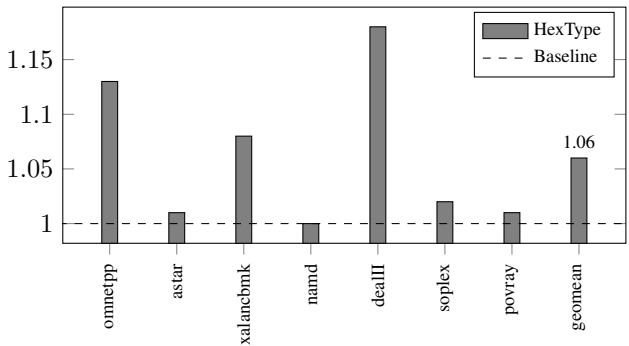
\pgfplotstableread[col sep=comma]{hextype.csv}{\hextype}
\pgfplotstablegetelem{0}{overhead}\of{\hextype}
\pgfmathsetmacro{\gm}{\pgfplotsretval}

\begin{figure}[H]
  \centering
  \begin{tikzpicture}
    \begin{axis}[
      ybar, bar width=6pt, legend cell align=left,
      legend style={font=\scriptsize}, area legend,
      height=0.2\textheight, width=0.5\textwidth,
      symbolic x coords={left,omnetpp,astar,xalancbmk,namd,dealII,soplex,povray,geomean,right},
      x tick label style={font=\scriptsize,rotate=90},
      xtick pos=left,
      xtick=data,
      ]
      \addplot[fill=gray] table[x={benchmark}, y={overhead}] {\hextype};
      \addplot[black,line legend,sharp plot,update limits=false,style=dashed]
      coordinates {(left, 1.0) (right, 1.0)};
      \node[black, above, style={font=\scriptsize}] at (axis cs: geomean,\gm) {\gm};
      \legend{HexType, Baseline}
    \end{axis}
  \end{tikzpicture}
\end{figure}

\subsubsection{Clang CFI}

We used the official version of LLVM/Clang 6.0.0 to compile the baseline
binaries, and used the \code{-fsanitize=cfi} flag for that same compiler to
generate the Clang CFI binaries. Clang CFI inserts checks (i) for casts between
\C++ class types and (ii) for indirect calls and \C++ member function calls
(virtual and non-virtual).
We enabled \code{-fno-sanitize-trap} to print diagnostic information.

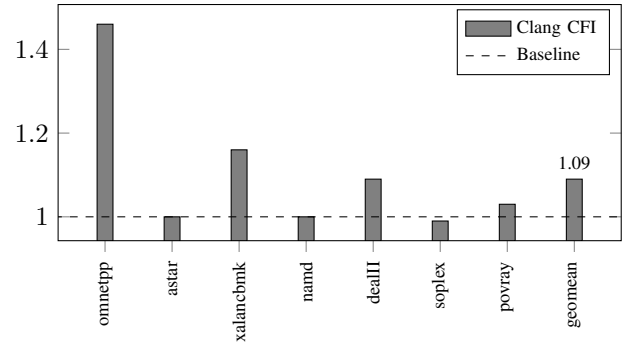
\pgfplotstableread[col sep=comma]{cfi.csv}{\cfi}
\pgfplotstablegetelem{0}{overhead}\of{\cfi}
\pgfmathsetmacro{\gm}{\pgfplotsretval}

\begin{figure}[H]
  \centering
  \begin{tikzpicture}
    \begin{axis}[
      ybar, bar width=6pt, legend cell align=left,
      legend style={font=\scriptsize}, area legend,
      height=0.2\textheight, width=0.5\textwidth,
      symbolic x coords={left,omnetpp,astar,xalancbmk,namd,dealII,soplex,povray,geomean,right},
      x tick label style={font=\scriptsize,rotate=90},
      xtick pos=left,
      xtick=data,
      ]
      \addplot[fill=gray] table[x={benchmark}, y={overhead}] {\cfi};
      \addplot[black,line legend,sharp plot,update limits=false,style=dashed]
      coordinates {(left, 1.0) (right, 1.0)};
      \node[black, above, style={font=\scriptsize}] at (axis cs: geomean,\gm) {\gm};
      \legend{Clang CFI, Baseline}
    \end{axis}
  \end{tikzpicture}
\end{figure}

\subsubsection{HexVASAN}

We applied HexVASAN's patches to the official version of LLVM/Clang 3.9.1 and
used that compiler to generate both the baseline and the HexVASAN binaries.  We
used the \code{-fsanitize=vasan} for the HexVASAN.  We disabled early program
termination after check failures to measure the run-time overhead for
\code{471.omnetpp}, which has a known false positive reported in the HexVASAN
paper.

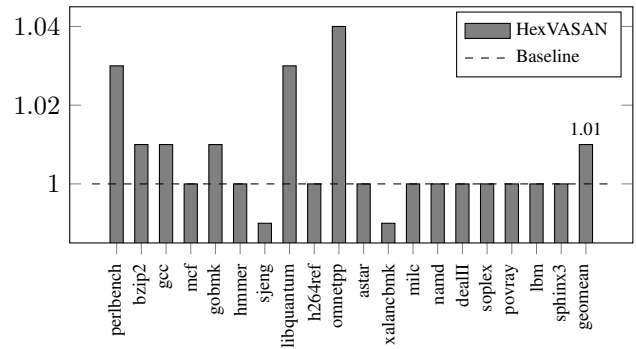
\pgfplotstableread[col sep=comma]{hexvasan.csv}{\hexvasan}
\pgfplotstablegetelem{0}{overhead}\of{\hexvasan}
\pgfmathsetmacro{\gm}{\pgfplotsretval}

\begin{figure}[H]
  \centering
  \begin{tikzpicture}
    \begin{axis}[
      ybar, bar width=5pt, legend cell align=left,
      legend style={font=\scriptsize}, area legend,
      height=0.2\textheight, width=0.5\textwidth,
      symbolic x coords={left,perlbench,bzip2,gcc,mcf,gobmk,hmmer,sjeng,libquantum,h264ref,omnetpp,astar,xalancbmk,milc,namd,dealII,soplex,povray,lbm,sphinx3,geomean,right},
      x tick label style={font=\scriptsize,rotate=90},
      xtick pos=left,
      xtick=data,
      ]
      \addplot[fill=gray] table[x={benchmark}, y={overhead}] {\hexvasan};
      \addplot[black,line legend,sharp plot,update limits=false,style=dashed]
      coordinates {(left, 1.0) (right, 1.0)};
      \node[black, above, style={font=\scriptsize}] at (axis cs: geomean,\gm) {\gm};
      \legend{HexVASAN, Baseline}
    \end{axis}
  \end{tikzpicture}
\end{figure}

\subsubsection{UndefinedBehaviorSanitizer}

We used the official version of LLVM/Clang 6.0.0 to compile the baseline
binaries, and used the \code{-fsanitize=undefined} flag for that same compiler
to generate the UBSan binaries. \code{-fsanitize=undefined} enables a carefully
chosen set of sanitizers. We refer the reader to the UBSan web page for a
description of the sanitizers in this set~\cite{ubsan}.

\pgfplotstableread[col sep=comma]{ubsan.csv}{\ubsan}
\pgfplotstablegetelem{0}{overhead}\of{\ubsan}
\pgfmathsetmacro{\gm}{\pgfplotsretval}

\begin{figure}[H]
  \centering
  \begin{tikzpicture}
    \begin{axis}[
      ybar, bar width=5pt, legend cell align=left,
      legend style={font=\scriptsize}, area legend,
      height=0.2\textheight, width=0.5\textwidth,
      symbolic x coords={left,perlbench,bzip2,gcc,mcf,gobmk,hmmer,sjeng,libquantum,h264ref,omnetpp,astar,xalancbmk,milc,namd,dealII,soplex,povray,lbm,sphinx3,geomean,right},
      x tick label style={font=\scriptsize,rotate=90},
      xtick pos=left,
      xtick=data,
      ]
      \addplot[fill=gray] table[x={benchmark}, y={overhead}] {\ubsan};
      \addplot[black,line legend,sharp plot,update limits=false,style=dashed]
      coordinates {(left, 1.0) (right, 1.0)};
      \node[black, above, style={font=\scriptsize}] at (axis cs: geomean,\gm) {\gm};
      \legend{UBSan, Baseline}
    \end{axis}
  \end{tikzpicture}
\end{figure}

\end{appendices}

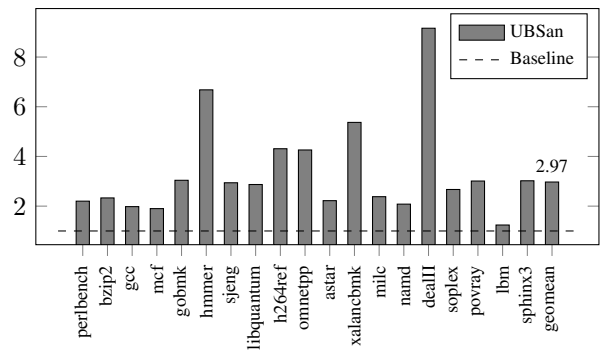
\end{document}